\documentclass[aps,pra,groupedaddress,tightenlines]{revtex4}

\usepackage{amsmath}
\usepackage{graphicx}
\usepackage[english]{babel}
\usepackage{hyperref}

\begin{document}

\title{Continuous-time quantum error correction\footnote{A chapter in the book \textit{Quantum Error Correction}, edited by Daniel A. Lidar and Todd A. Brun, (Cambridge University Press, 2013), 
\href{http://www.cambridge.org/us/academic/subjects/physics/quantum-physics-quantum-information-and-quantum-computation/quantum-error-correction}{http://www.cambridge.org/us/academic/subjects/physics/quantum-physics-quantum-information-and-quantum-computation/quantum-error-correction}.}}
\author{Ognyan Oreshkov}
\affiliation{QuIC, Ecole Polytechnique, CP 165, Universit\'{e} Libre de Bruxelles, 1050 Brussels, Belgium.}

\begin{abstract}
Continuous-time quantum error correction (CTQEC) is an approach to protecting quantum information from noise in which both the noise and the error correcting operations are treated as processes that are continuous in time.
This chapter investigates CTQEC based on continuous weak measurements and feedback
from the point of view of the subsystem principle, which states that protected quantum information is contained in a subsystem of the Hilbert space. We study how to approach the problem of constructing CTQEC protocols by looking at the
evolution of the state of the system in an encoded basis in which
the subsystem containing the protected information is explicit. This point of view allows us to reduce the problem to that of
protecting a known state, and to design CTQEC procedures
from protocols for the protection of a single qubit. We show
how previously studied CTQEC schemes with both direct and indirect feedback can be obtained from strategies for the protection
of a single qubit via weak measurements and weak unitary
operations. We also review results on the performance of CTQEC with direct feedback in cases of Markovian and non-Markovian
decoherence, where we have shown that due to the existence of a Zeno regime
in non-Markovian dynamics, the performance of CTQEC can exhibit a
quadratic improvement if the time resolution of the weak
error-correcting operations is high enough to reveal the
non-Markovian character of the noise process.
\end{abstract}

\maketitle



\section{Introduction}\label{introduction}

In the standard theory of quantum error correction, both the noise
and the error-correcting operations are represented by discrete
transformations. If $\mathcal{B}(\mathcal{H})$ denotes the space of
bounded operators over a Hilbert space $\mathcal{H}$, and
$\mathcal{H^S}$ is the (finite-dimensional) Hilbert space of the
controlled system, we say that the code subsystem $\mathcal{H}^A$ in
the decomposition
\begin{equation}
\mathcal{H^S}=\mathcal{H^A}\otimes\mathcal{H}^B\oplus
\mathcal{K}\label{decomposition}
\end{equation}
is correctable under the completely positive trace-preserving (CPTP)
noise map $\mathcal{E}:\mathcal{B}(\mathcal{H}^S)\rightarrow
\mathcal{B}(\mathcal{H}^S)$, if there exists a CPTP error-correcting
map $\mathcal{R}:\mathcal{B}(\mathcal{H}^S)\rightarrow
\mathcal{B}(\mathcal{H}^S)$, such that
\begin{gather}
\textrm{Tr}_B\{(\mathcal{P}^{AB}\circ\mathcal{R}\circ\mathcal{E})(\sigma)\}=\textrm{Tr}_B\{\sigma\},\label{correctablesystem}\\
\hspace{0.1cm} \textrm{for all }\sigma\in
\mathcal{B}(\mathcal{H}^S),
\sigma=\mathcal{P}^{AB}(\sigma)\hspace{0.1cm} ,\notag
\end{gather}
where $\mathcal{P}^{AB}(\cdot)$ denotes the superoperator projector
on $\mathcal{B}(\mathcal{H^A}\otimes\mathcal{H}^B)$.

This formalism is fundamental for the understanding of preserved
information under CPTP dynamics, but it depicts an idealized version
of the error-correction process. It represents both the noise and
the error-correcting operations as discrete CPTP maps, and assumes
that error correction is applied \textit{after} the noise. Such a
picture is a good approximation for the case when we are concerned
with error correction at a single instant via an operation which is
fast on the time scale of the noise, or in the case of repeated
error correction with fast operations in a regime where the
accumulation of uncorrectable errors can be ignored. In general,
however, a full error-correcting operation takes a finite time
interval during which the noise process is on. Furthermore, even if
we assume that error-correcting operations are instantaneous,
deviations from perfect correctability between repeated corrections
are unavoidable in any real situation. Thus in the case of
non-Markovian dynamics, the system may develop correlations with the
environment and the effective error maps between successive
corrections need not be completely positive. Therefore, a complete
description must take into account the continuous nature of both the
decoherence and the error-correction processes. Situations in which
both these processes are regarded as continuous in time are the
subject of continuous-time quantum error correction (CTQEC).

The first CTQEC model was proposed by Paz and Zurek (PZ)
\cite{Paz:1998:355} as a method of studying the performance of
repeated error correction with fast operations in the presence of
Markovian decoherence. Rather than describing the overall evolution
as a continuous decoherence process interrupted by instantaneous
error-correcting operations at discrete intervals, the authors
proposed to model the error-correcting procedure as a continuous
quantum-jump process, which allows a description of the evolution of
the system in terms of a continuous master equation in the Lindblad
form \cite{Lindblad:1976:119}. In this model, the infinitesimal error-correcting
transformation that the density matrix of the controlled system
undergoes during a time step $dt$ is
\begin{equation}\label{basicequation}
\rho\rightarrow (1-\kappa dt)\rho + \kappa dt \mathcal{R}(\rho),
\end{equation}
where $\mathcal{R}(\rho)$ is the completely positive trace-preserving
(CPTP) map describing a full error-correcting operation, and
$\kappa$ is the error-correction rate. The full error-correcting
operation $\mathcal{R}(\rho)$ can be thought of as consisting of a syndrome
detection, followed (if necessary) by a unitary operation
conditioned on the syndrome. The master equation describing the
evolution of a system subject to Markovian decoherence plus error
correction is then
\begin{equation}
\frac{d\rho}{dt}=\mathcal{L}(\rho)+\kappa\mathcal{J}(\rho),\label{errorcorrectionequation}
\end{equation}
where $\mathcal{L}(\rho)$ is the Lindblad generator describing the noise process, and
\begin{gather}
\mathcal{J}(\rho)=\mathcal{R}(\rho)-\rho
\end{gather}
is the quantum-jump error-correction generator. The Lindblad generator has the form
\begin{equation}
\mathcal{L}(\rho)=-i[H,\rho]+\frac{1}{2}\underset{j}{\sum}\lambda_j(2L_j\rho
L_j^{\dagger}-L_j^{\dagger}L_j\rho-\rho
L_j^{\dagger}L_j),\label{firstLindblad}
\end{equation}
where $H$ is a system Hamiltonian and the $\{L_j\}$ are suitably
normalized Lindblad operators describing different error channels
with decoherence rates $\lambda_j$. For example, the Lindbladian
\begin{equation}
\mathcal{L}(\rho)= \underset{j}{\sum}\lambda_j(X_j\rho X_j -
\rho),\label{Lbitflip}
\end{equation}
where $X_j$ denotes a local bit-flip operator acting the
$j^{\textrm{th}}$ qubit, describes independent Markovian bit-flip
errors.

The quantum-jump model can be viewed as a smoothed version of the
discrete scenario of repeated error correction, in which
instantaneous full error-correcting operations are applied at random
times with rate $\kappa$. It can also be looked upon as arising from
a continuous sequence of infinitesimal CPTP maps of the type
\eqref{basicequation}. In practice, such a weak map is never truly
infinitesimal, but rather has the form
\begin{equation} \rho
\rightarrow (1-\epsilon^2)\rho + \epsilon^2 \mathcal{R}(\rho),\label{wm}
\end{equation}
where $\epsilon \ll 1$ is a small but finite parameter, and the weak
operation takes a small but finite time $\tau_c$. For times $t$
much greater than $\tau_c$, the weak
error-correcting map (\ref{wm}) is well approximated by the
infinitesimal form \eqref{basicequation}, where the rate of error
correction is
\begin{equation}
\kappa = \epsilon^2 /\tau_c. \label{tauc}
\end{equation}
A weak map of the form \eqref{wm} could be implemented, for example,
by a weak coupling between the system and an ancilla via an
appropriate Hamiltonian, followed by discarding of the ancilla. The
continuous process in such a case corresponds to coupling the system
to a stream of fresh ancillas which continuously pump out the
entropy accumulated due to correctable errors. A closely related
scenario, where the ancilla is continuously cooled in order to reset
it to its initial state, was studied by Sarovar and Milburn in
Ref.~\cite{Sarovar:2005:012306}. Another possible implementation of the above
scheme is via weak measurements and weak unitary operations, as we
will see in this chapter.

If the set of errors $\{L_j\}$ are correctable by the code, the effect of the described CTQEC procedure is to slow down the rate
at which information is lost, and in
the limit of infinite error-correction rate (strong error-correcting
operations applied continuously often) the state of the system
freezes and is protected from errors at all times \cite{Paz:1998:355}. The effect of
freezing can be understood by noticing that the transformation
arising from decoherence during a short time step $\Delta t$, is
\begin{equation}
\rho\rightarrow \rho + \mathcal{L}(\rho)\Delta t +\textit{O}(\Delta t^2),
\end{equation}
i.e., the weight of correctable errors emerging during this time
interval is proportional to $\Delta t$, whereas uncorrectable errors
(higher-order terms) are of order $\textit{O}(\Delta t^2)$. Thus, if
errors are constantly corrected, in the limit $\Delta t \rightarrow
0$ uncorrectable errors cannot accumulate and the evolution stops.

The idea of using continuous weak operations for error correction
was developed further by Ahn, Doherty and Landahl (ADL) who proposed
a scheme for CTQEC based on continuous measurements of the error
syndromes and feedback operations conditioned on the measurement
record \cite{Ahn:2002:042301}. A continuous measurement is one
resulting from the continuous application of weak measurements,
i.e., measurements whose outcomes change the state by a small amount
\cite{Aharonov:1988:1351, Leggett:1989:2325, Peres:1989:2326,
Aharonov:1989:2327, Aharonov:1990:11, Brun:2002:719,
Oreshkov:2005:110409, Oreshkov:0812.4682}. As shown in
Refs.~\cite{Oreshkov:2005:110409, Oreshkov:0812.4682}, weak
measurements can be used to generate any quantum operation and
therefore provide a natural tool for approaching the problem of
error correction in continuous time. In the ADL scheme, the
evolution of the density matrix of the system subject to Markovian
noise with Lindbladian $\mathcal{L}$ and continuous-time quantum
error correction is described by the stochastic differential
equation
\begin{gather}
d\rho(t)=\mathcal{L}(\rho(t))dt+\frac{\kappa}{4}\sum_l\mathcal{D}[M_l](\rho(t))dt+\frac{\sqrt{\kappa}}{2}\sum_l\mathcal{F}[M_l](\rho(t))dW_l(t)\notag\\
-i\sum_r\lambda_r(\rho(t))[H_r,\rho(t)]dt,\label{ADL}
\end{gather}
where $\mathcal{D}[A](\rho) = A\rho A^{\dagger} -
\frac{1}{2}(A^{\dagger}A\rho+ \rho A^{\dagger}A)$,
$\mathcal{F}[A](\rho) = A\rho + \rho A^{\dagger} - \rho
\textrm{Tr}[A\rho + \rho A^{\dagger}]$, $M_l$ are the stabilizer
generators of the code, $W_l$ are Wiener processes (see
Sec.~\eqref{indirectsq}), and $H_r$ are correcting Hamiltonians that
are turned on with strength $\lambda_r(\rho)$ dependent on the state
of the system. Note that the encoded information is in principle
unknown, but the feedback is not conditioned on properties of the
state related to the encoded information. Thus in order to estimate
the state of the system at the present moment for the purpose of
applying feedback, one can assume that the encoded state was
initially the maximally mixed state. The parameters
$\lambda_r(\rho)$ are chosen so as to maximize the instantaneous
increase of the code-space fidelity, and are given by
$\lambda_r(\rho)=\lambda\textrm{sgn}\textrm{Tr}([\Pi_c,H_r]\rho)$,
where $\lambda$ is the maximum strength of the control Hamiltonians
and $\Pi_c$ is the projector on the code subspace. (Here the code is
assumed to be a standard stabilizer code.)

Following the ADL scheme, a number of variations of this approach
were proposed (see, e.g., Refs.~\cite{Ahn:2003:052310,
Sarovar:2004:052324,Chase:2008:032304}). All these schemes are to a
large extent heuristic, and their workings are not thoroughly
understood. The difficulty in rigorously motivating the construction
of error-correction protocols based on weak measurements and
feedback is that stochastic evolutions are generally too complicated
to study analytically. This is further complicated by the large
dimension of the Hilbert space of all qubits participating in the
code (note that even the problem of controlling a single qubit
generally requires numerical treatment \cite{Jacobs:2004:355}).
However, numerical simulations have shown that these schemes often
lead to a better performance in the presence of continuous noise
than the application of strong operations at finite time intervals.
Therefore, the use of continuous measurements and feedback seems to
offer a promising tool for decoherence control.

In this chapter, we will try to understand CTQEC and how to approach
the problem of constructing CTQEC protocols by looking at the
evolution of the state of the system in an encoded basis in which
the subsystem containing the protected information is explicit. We
will see that this point of view reduces the problem to that of
protecting a known state, and allows for designing CTQEC procedures
from protocols for the protection of a single qubit. We will show
how the PZ quantum-jump model and the ADL and similar schemes with
indirect feedback can be obtained from strategies for the protection
of a single qubit based on weak measurements and weak unitary
operations. We will also study the performance of CTQEC of the
quantum-jump type in the case of Markovian and non-Markovian
decoherence. We will show that due to the existence of a Zeno regime
in non-Markovian dynamics, the performance of CTQEC can exhibit a
quadratic improvement if the time resolution of the weak
error-correcting operations is sufficiently high to reveal the
non-Markovian character of the noise process.

\section{CTQEC in an encoded
basis}\label{sectionENCODEDBASIS}

As discussed in Chapter~6, correctable information is always
contained in subsystems of the system's Hilbert space
\cite{Knill:2006:042301, Blume-Kohout:2008:030501}. This means, in
particular, that if the information initially encoded in the
subsystem $\mathcal{H}^A$ in Eq.~\eqref{decomposition} is
correctable after the noise map $\mathcal{E}$, it is
\textit{unitarily recoverable} \cite{Kribs:2006:042329}, i.e., there
exists a unitary map $\mathcal{U}(\cdot)=U(\cdot)U^{\dagger}$,
$U\in\mathcal{B}(\mathcal{H}^S)$, such that
\begin{gather}
\mathcal{U}\circ\mathcal{E}(\rho\otimes\tau)=\rho\otimes\tau^{\prime },%
\hspace{0.4cm} \tau^{\prime }\in\mathcal{B}(\mathcal{H}^{B^{\prime
}}),
\label{unitarilyrecoverable2} \\
\hspace{0.2cm} \text{for all }\rho\in \mathcal{B}(\mathcal{H}^A), \hspace{%
0.1cm} \tau\in \mathcal{B}(\mathcal{H}^B),  \notag
\end{gather}
where the subsystem $\mathcal{H}^{B'}$ can be different from
$\mathcal{H}^{B}$. Complete correction generally requires an
additional CPTP map that transforms the operators on
$\mathcal{H}^{B^{\prime }}$ into operators on $\mathcal{H}^{B}$. As
shown in Ref.~\cite{Oreshkov:2008:022333},
Eq.~\eqref{unitarilyrecoverable2} is equivalent to the condition
that the Kraus operators $M_{\alpha}$ of $\mathcal{E}$ satisfy
\begin{gather}
M_{\alpha}P^{AB}=U^{\dagger}I^{A}\otimes C^{B\rightarrow
B'}_{\alpha},\hspace{0.2cm}C^{B\rightarrow B'}_{\alpha}:
\mathcal{H}^B\rightarrow \mathcal{H}^{B'}, \hspace{0.2cm} \forall
\alpha. \label{conditionKraus1}
\end{gather}
Observe that if a particular set of error operators $\{M_{i}\}$ is
correctable by the code, that is, if any CPTP map whose Kraus
operators are linear combinations of $\{M_{i}\}$ is correctable,
then there is a common recovery unitary $U$ for all such CPTP maps.
Note also that if the identity is among the correctable errors for
which the code is designed (this is the case, in particular, for all
stabilizer codes), from condition \eqref{conditionKraus1} it follows
that the unitary ${U}$ must leave the subsystem $\mathcal{H}^A$ in
$\mathcal{H}^A\otimes\mathcal{H}^B$ invariant up to a transformation
of the co-subsystem, $\mathcal{H}^B\rightarrow
\mathcal{H}^{\tilde{B}}$
($\textrm{dim}\mathcal{H}^{\tilde{B}}=\textrm{dim}\mathcal{H}^B$).
This means that if we change the basis by the unitary map ${U}$, the
effect of the error operators $M'_{\alpha}=UM_{\alpha}U^{\dagger}$
in the new basis is
\begin{gather}
M'_{\alpha}P^{A\tilde{B}}=UM_{\alpha}P^{AB}U^{\dagger}=I^{A}\otimes
C^{\tilde{B}\rightarrow
B'}_{\alpha},\hspace{0.2cm}C^{\tilde{B}\rightarrow B'}_{\alpha}:
\mathcal{H}^{\tilde{B}}\rightarrow \mathcal{H}^{B'}, \hspace{0.2cm}
\forall \alpha, \label{conditionKraus}
\end{gather}
i.e., the errors leave the code subsystem invariant up to a
transformation of the co-subsystem. A method of obtaining $U$ can be
found in Ref.~\cite{Kribs:2006:042329}.

In what follows, we will imagine for concreteness the case of an
$[[n,1,r,d]]$ operator stabilizer code. This is a code that encodes
$1$ qubit into $n$, has $r$ gauge qubits, and has distance $d$. In
the encoded basis defined above, the Hilbert space of all $n$ qubits
can be written as
\begin{gather}
\mathcal{H}^S=\mathcal{H}^A\otimes\bigotimes_{i=1}^{n-r-1}\mathcal{H}^s_i
\otimes\bigotimes_{j=1}^{r}\mathcal{H}^g_j,\label{decomp}
\end{gather}
where $\mathcal{H}^A$ is a subsystem which corresponds to the
logical qubit, $\mathcal{H}^s_i$ are the subsystems of the
\textit{syndrome} qubits, and $\mathcal{H}^g_j$ are the subsystems
of the \textit{gauge} qubits. Up to a redefinition of the basis of
the syndrome qubits, we can assume that the subspace
$\mathcal{H}^A\otimes\mathcal{H}^B$ in Eq.~\eqref{decomposition}
corresponds to
\begin{gather}
\mathcal{H}^A\otimes\mathcal{H}^B=\mathcal{H}^A\otimes\bigotimes_{i=1}^{n-r-1}|0\rangle^s_i
\otimes\bigotimes_{j=1}^{r}\mathcal{H}^g_j.\label{codespace}
\end{gather}
We will refer to this subspace loosely as the \textit{code space},
since this is where the state of the system is initialized, but we
must keep in mind that the information of interest is contained in
the tensor factor $\mathcal{H}^A$ in Eq.~\eqref{decomp}. If each of
the syndrome qubits is initialized in the state $|0\rangle$, any
correctable error will leave the subsystem $\mathcal{H}^A$ invariant
and will only affect the co-subsystem, most generally transforming
density operators on $\bigotimes_{i=1}^{n-r-1}|0\rangle^s_i
\otimes\bigotimes_{j=1}^{r}\mathcal{H}^g_j$ into density operators on
$\bigotimes_{i=1}^{n-r-1}\mathcal{H}^s_i
\otimes\bigotimes_{j=1}^{r}\mathcal{H}^g_j$. In this basis, an
error-correcting operation is simply a map on the syndrome qubits,
which returns them to the state $|00...0\rangle$. In the language of
stabilizer codes, a measurement of the syndrome is a measurement of
the state of all syndrome qubits in the $\{|0\rangle, |1\rangle\}$
basis, and a correcting operation is any operation that effectively
realizes a bit flip to those qubits which are in the state
$|1\rangle$.

If the syndrome qubits are not properly initialized (as for example,
after the occurrence of an error), a subsequent error generally
would not leave the code subsystem invariant. Most generally, after
a system subject to decoherence and error correction evolves for a
given time $t$, the state of the system becomes
\begin{gather}
\rho^S(t)=\alpha(t)\rho^{Ag}(t)\otimes\bigotimes_{i=1}^{n-r-1}|0\rangle\langle
0|^{s}_i+(1-\alpha(t))\widetilde{\rho}^{Ags}(t)+\textrm{cross
terms}.\label{densmat}
\end{gather}
Here $\rho^{Ag}(t)$ is a density matrix on the Hilbert space
$\mathcal{H}^A\otimes\bigotimes_{j=1}^{r}\mathcal{H}^g_j$,
$\widetilde{\rho}^{Ags}(t)$ is a density matrix with support on the
orthogonal complement $\widetilde{\mathcal{H}}$ of the code space
($\mathcal{H}^A\otimes\bigotimes_{i=1}^{n-r-1}|0\rangle^s_i
\bigotimes_{j=1}^{r}\mathcal{H}^g_j\oplus
\widetilde{\mathcal{H}}=\mathcal{H}^S$), $\alpha(t)\in[0,1]$ is the
code-space fidelity, and ``cross terms'' refers to linear
combinations of terms of the form
$|\psi_i\rangle\langle\widetilde{\phi}_j|$ and
$|\widetilde{\phi}_j\rangle\langle\psi_i|$, where
$\{|\psi_i\rangle\}$ is an orthonormal basis of
$\mathcal{H}^A\otimes\bigotimes_{i=1}^{n-r-1}|0\rangle^s_i
\otimes\bigotimes_{j=1}^{r}\mathcal{H}^g_j$ and
$\{|\widetilde{\phi}_i\rangle\}$ is an orthonormal basis of
$\widetilde{\mathcal{H}}$. The density matrix of the logical
subsystem is
$\mathcal{\rho}^A(t)=\alpha(t)\textrm{Tr}_g(\rho^{Ag}(t))+(1-\alpha(t))\textrm{Tr}_{gs}(\widetilde{\rho}^{Ags}(t))$,
where $\textrm{Tr}_g$ denotes partial tracing over the gauge qubits
and $\textrm{Tr}_{gs}$ denotes partial tracing over the gauge qubits
and the syndrome qubits. This density matrix is a transformed
version of the state initially encoded in the code subsystem, where
the transformation is the result of accumulation of uncorrectable
errors. (Note that any transformation inside the subsystem
$\mathcal{H}^A$ in Eq.~\eqref{decomp} is by definition
uncorrectable.)

Let us see how the density matrix $\rho^A$ changes as a result of
the action of the generator of noise during a time step $\Delta t$.
Since by assumption the action of the noise generator leaves the
code subsystem invariant up to a transformation of the co-subsystem,
its effect on the term
$\alpha\rho^{Ag}(t)\otimes\bigotimes_{i=1}^{n-r-1}|0\rangle\langle
0|^{s}_i$ in Eq.\eqref{densmat} during a time step $\Delta t$ does
not give rise to a non-trivial change in $\rho^A(t)$, but only to a
decrease in the code-space fidelity,
\begin{gather}
\alpha(t)\rightarrow \alpha(t)-\gamma(t)\alpha(t)\Delta t + O(\Delta
t^2),
\end{gather}
where $\gamma(t)\geq 0$ is a parameter which depends on the
characteristics of the noise process, such as the rates of different
errors, and possibly on the current density matrix of the gauge
qubits inside the code space, $\textrm{Tr}_A(\rho^{Ag}(t))$. Note
that if the noise is non-Markovian, the leading-order correction to
$\alpha(t)$ due to the action of the noise on
$\alpha\rho^{Ag}(t)\otimes\bigotimes_{i=1}^{n-r-1}|0\rangle\langle
0|^{s}_i$ is $O(\Delta t^2)$, i.e., $\gamma(t)=0$ (see
Sec.~\ref{nM}). The only way errors can arise inside the subsystem
$\mathcal{H}^A$ is by the action of the noise mechanism on the other
terms in Eq.~\eqref{densmat}. The weight of the second term is
$(1-\alpha(t))$, and during a single time step the noise generator
can give rise to a change in $\rho^A(t)$
\begin{gather}
\rho^A(t)\rightarrow \rho^A(t)+\delta \rho^A(t),
\end{gather}
where
\begin{gather}
\parallel\delta \rho^A(t)\parallel \leq B(1-\alpha(t))\Delta t+O(\Delta t^2), \hspace{0.2cm}
B\geq 0.\end{gather} The constant $B$ depends on the rate of the
noise process, its characteristics and the characteristics of the
code. From the positivity of the density matrix $\rho^S$ one can
show that the coefficients in front of the cross terms
$|\psi_i\rangle\langle\widetilde{\phi}_j|$ and
$|\widetilde{\phi}_j\rangle\langle\psi_i|$ are at most
$\sqrt{\alpha(1-\alpha)}$ in magnitude, and therefore the change
that can result in $\mathcal{\rho}^A$ due to the action of the noise
generator on the third term in Eq.\eqref{densmat} is limited by
\begin{gather}
\parallel\delta \rho^A(t)\parallel \leq C\sqrt{\alpha(1-\alpha(t))}\Delta t+O(\Delta t^2),
\end{gather}
where $C\geq 0$ is another constant dependent on the characteristics
of the noise and the code. Thus we see that the rate of change of
the density matrix $\rho^A$ is upper bounded as follows:
\begin{gather}
\parallel \frac{d\rho^A}{dt}\parallel \leq B(1-\alpha(t))+C\sqrt{\alpha(1-\alpha(t))}.
\end{gather}
In other words, if we manage to keep $(1-\alpha(t))$ small, we will
suppress the rate of accumulation of uncorrectable errors. The goal
of continuous-time quantum error correction can thus be understood
as that of keeping the state of every syndrome qubit close to the
state $|0\rangle$.

Notice that a strong error-correcting operation in this basis can be
realized by bringing each of the syndrome qubits to the state
$|0\rangle$ independently. Therefore, the problem of implementing a
strong error-correcting operation in terms of weak operations can be
reduced to the problem of implementing the corresponding
single-qubit operations via weak single-qubit operations. Of course,
this is not the most general way of realizing collective
initialization of the syndrome qubits, but it is appealing because
it reduces the task to that of addressing several independent qubits
individually. We will see, however, that the performance can be
enhanced if instead of addressing each of the syndrome qubits
individually, we address each syndrome which can be associated with
a qubit subspace in the space of the syndrome qubits. This will be
discussed in the next section. Here we note that the operations in
the original basis can be obtained by applying the inverse of the
basis transformation to the operations in the encoded basis.

To get an idea of what the transformation between bases looks like,
let us consider as an example the three-qubit bit-flip code with
stabilizer generated by $\{IZZ, ZZI \}$. This code has logical
codewords $|0_L\rangle=|000\rangle$ and $|1_L\rangle=|111\rangle$
and even though it only corrects bit-flip errors and does not have
gauge qubits, it captures all the characteristics of non-trivial
codes which are pertinent to our discussion. It can be verified that
a correcting unitary for this code is $U=U_cCX_{1,2}CX_{1,3}$, where
\begin{gather}
U_c=X_1\otimes|11\rangle\langle 11|_{23} + I_1\otimes (I_2\otimes
I_3-|11\rangle\langle 11|_{23}),\label{Uc}
\end{gather}
and $CX_{i,j}$ denotes the ``controlled not'' with qubit $i$ being
the control and qubit $j$ the target. This unitary transforms the
single-qubit bit-flip error operators as
\begin{eqnarray}
XII&\rightarrow & I\otimes(|00\rangle\langle 11|+|11\rangle\langle
00|)+X\otimes(|01\rangle\langle 10|+|10\rangle\langle 01|),\notag\\
IXI&\rightarrow & I\otimes X\otimes |0\rangle\langle 0|+X\otimes
X\otimes |1\rangle\langle
1|,\notag\\
IIX&\rightarrow & I\otimes|0\rangle\langle 0|\otimes
X+X\otimes|1\rangle\langle 1|\otimes X.
\end{eqnarray}
In this basis, when the second and third qubits are in the state
$|0\rangle$, the error operators leave the state of the first qubit
invariant. Going back to the original basis is achieved by applying
the basis transformation backwards, i.e., by applying the unitary
$CX_{1,3}CX_{1,2}U_c$.

\section{Quantum-jump CTQEC with weak measurements}\label{sectionQJ}

\subsection{The single-qubit problem}\label{sqp}

In this section we will show how to implement the PZ quantum-jump
error-correction scheme (Eq.~\eqref{basicequation}) using weak
measurements in the encoded basis. We start with the problem of
protecting a single qubit in the state $|0\rangle$ from noise using
weak measurements. The state $|0\rangle$ can be thought of as a
trivial stabilizer code with stabilizer generated by $Z$. We will
first consider the case of Markovian bit-flip decoherence, since
this model is simple and provides a good intuition. Later, we will
extend the result to general noise models.

A Markovian bit-flip process is described by the master equation
\begin{equation}
\frac{d\rho(t)}{dt} = \gamma ( X \rho X - \rho).
\end{equation}
where $\gamma$ is the bit-flip rate. The general solution to this
equation is
\begin{equation}
\rho(t) =\frac{1+e^{-2\gamma t}}{2}\rho(0)+\frac{1-e^{-2\gamma
t}}{2}X\rho(0)X. \label{solu1}
\end{equation}
If the system starts in the state $|0\rangle\langle 0|$, without
error correction it will decay down the $Z$-axis towards the
maximally mixed state.

In the language of stabilizer codes, an error-correcting operation
for this code consists of a measurement of the stabilizer generator
$Z$ followed by a unitary correction. If the result is $|1\rangle$,
we apply a bit-flip operation $X$, and if the result is $|0\rangle$,
we do nothing. The completely positive map corresponding to this
strong error-correcting operation is
\begin{equation}
\mathcal{R}(\rho)= X |1\rangle \langle 1| \rho |1\rangle \langle 1| X +
|0\rangle \langle 0| \rho |0\rangle \langle 0| =|0\rangle \langle 1|
\rho |1\rangle \langle 0|+|0\rangle \langle 0| \rho |0\rangle
\langle 0|.\label{singlequbitstrongmap}
\end{equation}
One heuristic approach to making the above procedure continuous is
to consider weak measurements of the stabilizer generator $Z$ and
weak rotations around the $X$-axis of the Bloch sphere conditioned
on the measurement record. This is exactly the approach considered
in the feedback procedures of the ADL type, and we will discuss it
in Sec.~\ref{indirectsq}.

Observe that the transformation \eqref{singlequbitstrongmap} can
also be written as
\begin{gather}
\mathcal{R}(\rho)= |0\rangle \langle +| \rho |+\rangle \langle
0|+|0\rangle \langle -| \rho |-\rangle \langle 0|=\notag\\
 ZW |+\rangle
\langle +| \rho |+\rangle \langle +|WZ + XW|-\rangle \langle -| \rho
|-\rangle \langle -|WX,
\end{gather}
where $|\pm\rangle = (|0\rangle\pm |1\rangle)/\sqrt{2}$ and $W$ is
the Hadamard gate. Therefore the same error-correcting operation can
be implemented as a measurement in the $\{ |+\rangle, |-\rangle\}$ basis
(measurement of the operator $X$), followed by a unitary conditioned
on the outcome: if the outcome is $|+\rangle$, we apply $ZW$; if the
outcome is $|-\rangle$, we apply $XW$. This choice of unitaries is
not unique---for example, we could apply just $W$ instead of $ZW$
after outcome $|+\rangle$. But this particular choice has a
convenient geometric interpretation---the unitary $ZW$ corresponds
to a rotation around the $Y$-axis by an angle $\pi/2$, $ZW =
e^{i\frac{\pi}{2}\frac{Y}{2}}$, and $XW$ corresponds to a rotation
around the same axis by an angle $-\pi/2$, $ZW =
e^{-i\frac{\pi}{2}\frac{Y}{2}}$.

A weak version of the above error-correcting operation can be
constructed by taking the corresponding weak measurement of the
operator $X$, followed by a weak rotation around the $Y$-axis, whose
direction is conditioned on the outcome:
\begin{equation}
\begin{split}
 \rho \rightarrow \frac{I+i\epsilon'Y}{\sqrt
{1+{\epsilon'}^2}}\sqrt{\frac{I+\epsilon
X}{2}}\rho\sqrt{\frac{I+\epsilon
X}{2}}\frac{I-i\epsilon'Y}{\sqrt {1+{\epsilon'}^2}}+ \\
 +\frac{I-i\epsilon'Y}{\sqrt
{1+{\epsilon'}^2}}\sqrt{\frac{I-\epsilon
X}{2}}\rho\sqrt{\frac{I-\epsilon X}{2}}\frac{I+i\epsilon'Y}{\sqrt
{1+{\epsilon'}^2}}.\label{singlequbitweakmap}
\end{split}
\end{equation}
Here $\epsilon$ and $\epsilon'$ are small parameters. Note that the
fact that we describe the net result of the transformation by a CPTP
map means that after we apply feedback, we discard information about
the outcome of the measurement, or rather, we do not condition any
future operations on that information and therefore the
transformation of the average density matrix during a single time
step is given by Eq.~\eqref{singlequbitweakmap}. Such a scheme is
said to be based on \textit{direct} feedback, i.e., the feedback
Hamiltonian depends only on the outcome of the most recent
measurement, which does not require information processing of the
measurement record. Generally, discarding information leads to
suboptimal protocols, and we will discuss the possibility of
improving that scheme in Sec.~\ref{indirectsq}.

From the symmetry of the map \eqref{singlequbitweakmap} it can be
seen that if the map is applied to a state which lies on the
$Z$-axis, it will keep the state on the $Z$-axis. Whether the state
will move towards $|0\rangle\langle 0|$ or towards $|1\rangle\langle
1|$, depends on the relation between $\epsilon$ and $\epsilon'$.
Since our goal is to protect the state from drifting away from
$|0\rangle\langle 0|$ due to bit-flip decoherence, for now we will
assume that the state lies on the $Z$-axis in the northern
hemisphere. We would like, if possible, to choose the relation
between the parameters $\epsilon$ and $\epsilon'$ in such a way that
the effect of this map on any state on the $Z$-axis to be to move
that state towards $|0\rangle\langle 0|$.

In order to calculate the effect of this map on a given state, it is
convenient to write the state in the $\{|+\rangle,|-\rangle\}$ basis. For a
state on the $Z$-axis, $\rho = \alpha |0\rangle\langle 0|+(1-\alpha)
|1\rangle\langle 1|$, we have
\begin{equation}
\rho = \frac{1}{2}|+\rangle\langle +|+ \frac{1}{2}|-\rangle\langle
-| + (2\alpha -1)\left(\frac{1}{2}|+\rangle\langle -|+
\frac{1}{2}|-\rangle\langle +|\right).\label{rhopm}
\end{equation}
For the action of our map on the state \eqref{rhopm} we obtain:
\begin{gather}
\rho \rightarrow \frac{1}{2}|+\rangle\langle +|+
\frac{1}{2}|-\rangle\langle -| \notag\\+
\frac{(1-{\epsilon'}^2)\sqrt{1-\epsilon^2}(2\alpha -1) +
2\epsilon\epsilon'}{1+{\epsilon'}^2}\left(\frac{1}{2}|+\rangle\langle
-|+ \frac{1}{2}|-\rangle\langle +|\right).\label{transf}
\end{gather}
Thus we can think that upon this transformation the parameter
$\alpha$ transforms to $\alpha'$, where
\begin{equation}
2\alpha'-1 =\frac{(1-{\epsilon'}^2)\sqrt{1-\epsilon^2}(2\alpha -1) +
2\epsilon\epsilon'}{1+{\epsilon'}^2}.\label{alpha'}
\end{equation}
If it is possible to choose the relation between $\epsilon$ and
$\epsilon'$ in such a way that $\alpha'\geq \alpha$ for every $0\leq
\alpha \leq 1$, then clearly the state must remain invariant when
$\alpha = 1$. Imposing this requirement, we obtain
\begin{equation}
\epsilon = \frac{2\epsilon'}{1+{\epsilon'}^2},
\end{equation}
or equivalently
\begin{equation}
\epsilon'=\frac{1-\sqrt{1-\epsilon^2}}{\epsilon}.\label{varepsilon'}
\end{equation}
Substituting back in \eqref{alpha'}, we can express
\begin{equation}
\alpha'-\alpha =
\frac{4{\epsilon'}^2}{(1+{\epsilon'}^2)^2}(1-\alpha)\geq 0.
\end{equation}
We see that the coefficient $\alpha$ (which is the fidelity of the
state with $|0\rangle\langle 0|$) indeed increases after every
application of our weak completely positive map. The amount by which
it increases for fixed $\epsilon'$ depends on $\alpha$ and becomes
smaller as $\alpha$ approaches 1.

Since we will be taking the limit $\epsilon\rightarrow 0$, we can
write Eq.~\eqref{varepsilon'} as
\begin{equation}
\epsilon'=\frac{\epsilon}{2}+\textit{O}(\epsilon^3).\label{epsilonprime}
\end{equation}
If we define the relation between the time step $\tau_c$ and
$\epsilon$ as in Eq.~\eqref{tauc}, for the effect of the CPTP map
\eqref{singlequbitweakmap} on an arbitrary state of the form $\rho =
\alpha|0\rangle \langle 0 | +\beta |0\rangle \langle 1| + \beta^*
|1\rangle \langle 0| + (1-\alpha)|1\rangle \langle 1|$, $\alpha\in
R$, $\beta\in C$, we obtain
\begin{gather}
\alpha \rightarrow \alpha + (1-\alpha) \kappa \tau_c,\\
\beta \rightarrow \sqrt{1-\kappa \tau_c} \beta = \beta -
\frac{1}{2}\kappa \beta \tau_c +
O({\tau_c}^2).\label{effectofmapgen}
\end{gather}
This is exactly the map \eqref{wm} for $\mathcal{R}(\rho)$ given by
Eq.~\eqref{singlequbitstrongmap}.

We see that for an infinitesimal time step $dt$, the effect of the
noise is to decrease $\alpha(t)$ by the amount $\lambda
(2\alpha(t)-1) dt$ and that of the correcting operation is to
increase it by $\kappa (1-\alpha(t)) dt$. Combining both effects, we
obtain the net master equation that describes the evolution of the
qubit subject to Markovian bit-flip errors and the quantum-jump
error-correction scheme:
\begin{equation} \label{equation1}
\frac{d\alpha(t)}{dt}=-(\kappa+2\lambda)\alpha(t)+(\kappa +
\lambda).
\end{equation}
The solution is
\begin{equation}
\alpha(t)=(1-\alpha_*)e^{-(\kappa+2\lambda)t}+\alpha_*, \label{MSQS}
\end{equation}
where
\begin{equation}
\alpha_*=1-\frac{1}{2+r}, \label{attractor}
\end{equation}
and $r=\kappa/\lambda$ is the ratio between the rate of error
correction and the rate of decoherence. We see that the fidelity
decays, but it is confined above its asymptotic value $\alpha_*$
which can be made arbitrarily close to 1 for sufficiently large $r$.

Finally, let us show that this procedure works for any kind of
decoherence where the state need not remain on the $Z$-axis at all
times. From Eq.~\eqref{effectofmapgen} we see that the effect of a
single application of the map to a general state is to transfer a
small portion of the $|1\rangle \langle 1|$-component to $|0\rangle
\langle 0|$, and to decrease the magnitude of the off-diagonal
components by multiplying them by $\sqrt{1-\kappa \tau_c}$. If there
is noise, the most general negative effect of a single step of the
noise process is to increase the magnitude of $\beta$ and decrease
$\alpha$. For a realistic physical map, the amounts by which these
components change during a time step $\Delta t$ should tend to zero
when $\Delta t \rightarrow 0$. Since ultimately any noise process is
driven by a Hamiltonian acting on the system and its environment,
this means that for small $\Delta t$, each of these amounts can be
upper-bounded by $\gamma_{max} \Delta t$, where $\gamma_{max}$ is
some finite positive number. Therefore, if the system is
simultaneously subject to decoherence and error correction,
$|\beta|$ and $(1-\alpha)$ will not increase above certain values
for which the single-step effects of decoherence and
error-correction exactly cancel each other. We can upper-bound these
quantities by
\begin{gather}
(1-\alpha)_{max} = \frac{\gamma_{max}}{\kappa},\\
|\beta|_{max} = \frac{2\gamma_{max}}{\kappa}.
\end{gather}
This means that the state can be kept arbitrarily close to
$|0\rangle \langle 0|$ for sufficiently high rates of error
correction $\kappa$. In Sec.~\ref{exact} we will see that if the
noise is non-Markovian, $(1-\alpha)_{max}$ scales as
$\frac{1}{\kappa^2}$ for large $\kappa$!

We remark that one way of implementing the weak measurement of the
$X$ operator used in this scheme, is by coupling the system qubit to
an ancilla qubit prepared in the state $|+\rangle\langle +|$ for a
short time, via the Hamiltonian $H_X=-X\otimes Y$ where $X$ acts on
the system qubit and $Y$ acts on the ancilla, followed by a
measurement of the ancilla in the $\{|0\rangle, |1\rangle\}$ basis
(the latter can be destructive). It can be verified that if we first
apply the unitary transformation
$U_X(\epsilon)=\exp{i\frac{\epsilon}{2}X\otimes Y}$ followed by a
measurement of the ancilla, up to second order in $\epsilon$ the
resulting measurement on the system is
\begin{gather}
\rho\rightarrow \frac{\sqrt{\frac{I\pm\epsilon
X}{2}}\rho\sqrt{\frac{I\pm\epsilon X}{2}}}{p_{\pm}},
\end{gather}
with probabilities
$p_{\pm}=\frac{1}{2}(1\pm\epsilon\textrm{Tr}(X\rho))$. Since we are
interested in the limit where $\epsilon\rightarrow 0$, only the
lowest-order nontrivial contributions to the error-correcting CPTP
map are important, and they are of order $\epsilon^2$.

\subsection{General codes}

How do we extend this approach to general codes? As we mentioned
earlier, one way is to simply apply the described operation to each
of the syndrome qubits in the encoded basis. According to the
argument in the previous subsection, no matter what the exact form
of the noise process on the syndrome qubits is, this scheme will
keep each of them close to the state $|0\rangle\langle 0|$ within
some distance that can be made arbitrarily small for sufficiently
large error-correction rates. This in turn would ensure that the
code-space fidelity is close to $1$, which would suppress the rate
of accumulation of uncorrectable errors as argued in
Sec.~\ref{sectionENCODEDBASIS}. This approach is particularly
attractive because of its conceptual simplicity and the fact that it
involves operations only on each of the syndrome qubits whose number
$n-r-1$ is smaller than the number of different nontrivial
correctable errors which can be up to $2^{n-r-1}-1$. Furthermore,
it is obvious that the operations on the different qubits commute
and therefore can be applied simultaneously. However, it is not
difficult to see that even though the equivalent infinitesimal map
has the form \eqref{basicequation}, the effective $\mathcal{R}(\rho)$ is
not equal to the error correcting map for this code, where the
latter acts as
\begin{gather}
\mathcal{R}(\rho^s)= \bigotimes_{i=1}^{n-r-1}|0\rangle\langle 0|^{s}_i
\label{Phiofrho}
\end{gather}
for any state $\rho^s$ of all syndrome qubits. This is because, if
we apply error correction separately on the different qubits, up to
first order in $dt$ only those terms in which there is one qubit in
the state $|1\rangle$ and all the rest are in the state $|0\rangle$
(such as, e.g., $|10...0\rangle\langle 10...0|$) will get mapped to
$|00...0\rangle\langle 00...0|$. The full error-correcting map,
however, maps all states to the state $|00...0\rangle\langle
00...0|$ and therefore it is more powerful. Is there a way to
construct the full map based on the single-qubit operations
described in the previous subsection?

It turns out that the answer is yes. The idea is to associate an
abstract qubit to each non-trivial error syndrome in the code as
follows. As was mentioned earlier, each syndrome corresponds to a
state of the syndrome qubits of the form
$|\nu_1\nu_2...\nu_{n-r-1}\rangle$, where $\nu_i$ can be either $0$
or $1$. Let us label these different syndrome states by
$|i_s\rangle$, $i_s=0,...,2^{n-r-1}-1$, with
$|0_s\rangle=|00...0\rangle$ being the trivial syndrome
corresponding to ``no error''. The density matrix of the entire
system can then be written
\begin{gather}
\rho^S=\alpha(t)\rho^{Ag}(t)\otimes |0_s\rangle\langle 0_s|
+\sum_{i_s\geq 1}\beta_{i_s}\widetilde{\rho}^{Ag}_{i_s}(t)\otimes
|i_s\rangle\langle i_s|\notag\\
+\sum_{i_s\neq j_s}\sigma^{Ag}_{i_sj_s}(t)\otimes |i_s\rangle
\langle j_s|,\label{dmsb}
\end{gather}
where $\widetilde{\rho}^{Ag}_{i_s}(t)$ are density matrices on
$\mathcal{H}^A\otimes\bigotimes_{j=1}^{r}\mathcal{H}^g_j$,
$\beta_{i_s}\geq 0$ are the weights of the state inside the
different error subspaces, and $\sigma^{Ag}_{i_sj_s}(t)$ are
operators on
$\mathcal{H}^A\otimes\bigotimes_{j=1}^{r}\mathcal{H}^g_j$.

To each nontrivial syndrome we can associate a qubit subspace of the
space of all syndrome qubits, which is spanned by the state
$|0_s\rangle$ and the state $|i_s\rangle$ corresponding to that
syndrome. Let us take for concreteness one of these qubits---the
subspace spanned by $|0_s\rangle$ and $|1_s\rangle$. If we apply the
single-qubit operations described in the previous subsection to this
subspace while acting trivially on its orthogonal complement, the
effect of the resulting operation on the terms
$\beta_{1_s}\widetilde{\rho}^{Ag}_{1_s}(t)\otimes |1_s\rangle\langle
1_s|+\sigma^{Ag}_{0_s1_s}(t)\otimes |0_s\rangle \langle
1_s|+\sigma^{Ag}_{1_s0_s}(t)\otimes |1_s\rangle \langle 0_s|$ in
Eq.~\eqref{dmsb} will be the same as that of the quantum-jump error
correcting map \eqref{basicequation} with $\mathcal{R}(\rho)$ given
by Eq.~\eqref{Phiofrho}. At the same time, the effect on the rest of
the terms will be trivial. Therefore, if we apply the analogous
operation to each of the qubit subspaces spanned by $|i_s\rangle$
and $|0_s\rangle$, we will effectively realize the desired
quantum-jump error correcting map.

Observe that all these single-qubit maps commute and so do the
generators they give rise to in the corresponding continuous
quantum-jump equation. If we think of the resulting processes as
being driven by the action of the quantum-jump generators, then it
is obvious that all of them can be implemented simultaneously.
However, if we think of each of these maps as resulting from weak
measurements and weak unitary operations as described in the
previous subsection, the measurements and unitaries do \textit{not}
commute. For example, the $X$ operator for the $j_s^{\textrm{th}}$
qubit has the form $X_{j_s}=|j_s\rangle\langle
0_s|+|0_s\rangle\langle j_s|$, and therefore
$[X_{i_s},X_{j_s}]=|i_s\rangle\langle j_s|-|j_s\rangle\langle i_s|$.
This means that the measurements of the $X$ operators cannot be
implemented simultaneously on all qubits. The same holds for the
rotations around the $Y$-axes. Does this mean that we have to apply
the different operations in series? This would require the ability
to precisely turn on and off, on a very short time scale, the
couplings to the external fields needed for the different
measurements, which does not correspond to a continuous measurement.

It turns out that alternating the different couplings is not
needed---the same couplings that one would use for implementing the
weak measurements on the individual qubits can be turned on
simultaneously, and so can the feedback Hamiltonians that one would
use depending on the outcomes of the different measurements. This is
because all extra terms that arise from the fact that the operations
on the different qubits do not commute, cancel out when we average
over the outcomes. We outline how this can be verified using the
implementation of the weak measurement via a qubit ancilla described
at the end of Sec.~\ref{sqp}. For each of the qubits corresponding
to different syndromes, we will need to turn on a different
Hamiltonian that couples that qubit to a separate ancilla initially
prepared in the state $|+\rangle$. Let us label the ancilla
corresponding to the ${j_s}^{\textrm{th}}$ qubit also by $j_s$. If
we turn on all of these Hamiltonians simultaneously, the overall
Hamiltonian is
\begin{gather}
H_{meas}=-\sum_{j_s}X_{j_s}\otimes Y^a_{j_s},
\end{gather}
where the $Y^a_{j_s}$ act on the different ancilla systems but the
$X_{j_s}$ do not act on different systems and do not commute.
Imagine that this Hamiltonian acts for time $\frac{\epsilon}{2}$,
i.e., it gives rise to the unitary
$U=\exp(\frac{i\epsilon}{2}\sum_{j_s}X_{j_s}\otimes Y^a_{j_s})$. At
this point we can measure projectively each of the ancillas in the
$\{|0\rangle, |1\rangle\}$ basis and turn on the corresponding
single-qubit correction Hamiltonians $\xi_{j_s}Y_{j_s}$ where
$\xi_{j_s}=\pm 1$ is the sign of the Hamiltonian which depends on
the outcome of the measurement, and
$Y_{j_s}=i|j_s\rangle\langle0_s|-i|0_s\rangle\langle j_s|$. The
overall feedback Hamiltonian is
\begin{gather}
H_{fb}=\sum_{j_s}\xi_{j_s}Y_{j_s}.
\end{gather}
One can verify that up to second order in $\epsilon$, the resulting
operation after averaging over the outcomes is exactly equal to the
quantum jump operation \eqref{basicequation} with $\mathcal{R}(\rho)$ given
by Eq.~\eqref{Phiofrho}. The easiest way to see this is to observe
that all unwanted terms in the resulting density matrix are
proportional to $\xi_{i_s}\xi_{j_s}$, $i_s\neq j_s$, and therefore
when we sum over all different outcomes, these terms disappear.

To get an idea of what the weak measurements and feedback unitaries
mean in the original basis, let us look again at the three-qubit
bit-flip code. Observe that the syndrome states $|i\rangle^s$ in the
encoded basis are $|1_s\rangle=|10\rangle$,
$|2_s\rangle=|01\rangle$, $|3_s\rangle=|11\rangle$, i.e., the three
abstract qubits corresponding to these syndromes have $X$ and $Y$
operators
\begin{eqnarray}
X_{1_s}&=&I\otimes (|10\rangle\langle 00|+|00\rangle\langle
10|+|01\rangle\langle 01|+|11\rangle\langle 11|),\notag\\
X_{2_s}&=&I\otimes (|01\rangle\langle 00|+|00\rangle\langle
01|+|10\rangle\langle 10|+|11\rangle\langle 11|),\notag\\
X_{3_s}&=&I\otimes (|11\rangle\langle 00|+|00\rangle\langle
11|+|01\rangle\langle 01|+|10\rangle\langle 10|),
\end{eqnarray}
\begin{eqnarray}
Y_{1_s}&=&I\otimes (i|10\rangle\langle 00|-i|00\rangle\langle
10|+|01\rangle\langle 01|+|11\rangle\langle 11|),\notag\\
Y_{2_s}&=&I\otimes (i|01\rangle\langle 00|-i|00\rangle\langle
01|+|10\rangle\langle 10|+|11\rangle\langle 11|),\notag\\
Y_{3_s}&=&I\otimes (i|11\rangle\langle 00|-i|00\rangle\langle
11|+|01\rangle\langle 01|+|10\rangle\langle 10|).
\end{eqnarray}
By applying the inverse basis transformation $CX_{1,3}CX_{1,2}U_c$
with $U_c$ given by Eq.~\eqref{Uc}, we obtain these operators in the
original basis:
\begin{gather}
X^{\prime}_{1_s}=\frac{1}{2}ZXZ+\frac{1}{2}IXI+\frac{1}{2}III-\frac{1}{2}ZIZ,\notag\\
X^{\prime}_{2_s}=\frac{1}{2}ZZX+\frac{1}{2}IIX+\frac{1}{2}III-\frac{1}{2}ZZI,\notag\\
X^{\prime}_{3_s}=\frac{1}{2}XZZ+\frac{1}{2}XII+\frac{1}{2}III-\frac{1}{2}IZZ,\label{Xsprime}
\end{gather}
\begin{gather}
Y^{\prime}_{1_s}=\frac{1}{2}ZYI+\frac{1}{2}IYZ+\frac{1}{2}III-\frac{1}{2}ZIZ,\notag\\
Y^{\prime}_{2_s}=\frac{1}{2}IZY+\frac{1}{2}ZIY+\frac{1}{2}III-\frac{1}{2}ZZI,\notag\\
Y^{\prime}_{3_s}=\frac{1}{2}YZI+\frac{1}{2}YIZ+\frac{1}{2}III-\frac{1}{2}IZZ.
\label{Ysprime}
\end{gather}
We see that implementing the PZ scheme using weak measurements and
unitary operations requires the ability to apply Hamiltonians which
are complicated sums of different elements of the Pauli group. We
will postpone the analysis of the performance of that scheme in the
presence of decoherence until Sec.~\ref{exact}. We now turn to look
at alternative methods for protecting a single qubit from noise
using weak measurements, and their corresponding generalizations to
multi-qubit codes.

\section{Schemes with indirect feedback}\label{indirect}

\subsection{The single-qubit problem}\label{indirectsq}

We already mentioned that another way of ``continuization'' of the
discrete single-qubit error-correcting map
\eqref{singlequbitstrongmap} is to apply continuous measurements of
the stabilizer generator $Z$ and rotations around the $X$-axis
conditioned on the measurement record. A continuous measurement of
the operator $Z$ can be achieved by an infinite repetition of a weak
measurement with measurement operators
\begin{equation}
M^Z_{\pm}(\epsilon)= \sqrt{\frac{I\pm\tanh(\epsilon) Z}{2}}.
\end{equation}
The evolution of the state of the system under such observation can
be described by a random walk along a curve parameterized by $x\in
R$. The state at any moment during the procedure can be written in
the form
\begin{equation}
\rho(x) =\frac{M^Z(x)\rho(0)M^Z(x)}{\textrm{Tr}(M^Z(x)\rho(0)
M^Z(x))}
\end{equation}
for some value of $x$, where $M^Z(x) = \sqrt{(I+\tanh(x) Z)/2}$ and
$\rho(0)$ is the initial state. After every application of the weak
measurement $M^Z_{\pm}(\epsilon)$, the parameter $x$ changes to
$x\pm\epsilon$ depending on the outcome. The two projective
measurement outcomes of the strong measurement of $Z$ correspond to
$x=\pm\infty$. The procedure is continued until $|x|\geq X$ for some
$X$ which is sufficiently large that $M^Z(X)\approx |0\rangle\langle
0|$ and $M^Z(-X)\approx |1\rangle\langle 1|$ to any desired
precision \cite{Oreshkov:0812.4682}.

In the limit when $\epsilon\rightarrow 0$, the evolution of the
state of the system can be described by a continuous stochastic
differential equation. We can introduce a time step $\delta t$ and a
rate
\begin{equation}
\kappa=\epsilon^2/\delta t.
\end{equation}
Then we can define a mean-zero increment $\delta W$ as follows:
\begin{gather}
\delta W=(\delta x-M[\delta x])/\sqrt{\kappa},
\end{gather}
where $\delta x=\pm\epsilon$ and $M[\delta x]$ is the mean of
$\delta x$,
\begin{equation}
M[\delta x]=\epsilon (p_+(x)-p_-(x)).
\end{equation}
Here $p_{\pm}(x)$ are the probabilities for the two outcomes of the
weak measurement $M^Z_{\pm}(\epsilon)$ at the point $x$,
\begin{equation}
p_{\pm}(x)=\frac{1}{2}(1\pm \epsilon\langle Z\rangle_x ),
\end{equation}
with $\langle Z\rangle_x=\textrm{Tr}(Z\rho(x))$. Note that
$M[(\delta W)^2]=\delta t+\textit{O}(\delta t^2)$.

Expanding the change of the state under the measurement
$M^Z_{\pm}(\epsilon)$ up to second order in $\delta W$, and taking
the limit $\delta W\rightarrow 0$ while keeping the rate $\kappa$
fixed, it can be shown that the evolution of the state of the system
subject to such a continuous observation is described by the
following stochastic differential equation:
\begin{equation}
d\rho(t) = \frac{\kappa}{4} \mathcal{D}[Z](\rho(t)) dt +
\frac{\sqrt{\kappa}}{2}\mathcal{F}[Z](\rho(t)) dW(t).
\end{equation}
Here $dW(t)$ is a Wiener increment, i.e., a mean-zero normally
distributed random variable with variance $dt$. The evolution of the
parameter $x$ is given by
\begin{equation}
dx(t)= \kappa \langle Z \rangle_t dt + \sqrt{\kappa} dW(t),
\end{equation}
where $\langle Z \rangle_t=\textrm{Tr}(Z\rho(t))$. From $x(t)$ one
can define the average measurement current as the mean of
$dx(t)/dt$,
\begin{gather}
I^{ave}_x(t)= \kappa \langle Z\rangle_t.\label{meascurr}
\end{gather}

If we apply no error correction to our qubit (initially in the state
$|0\rangle\langle 0|$), under bit-flip decoherence its state will
drift down the $Z$-axis of the Bloch sphere towards the center of
the sphere (the maximally mixed state). According to the scheme
proposed in Ref.~\cite{Ahn:2002:042301}, at a given moment we apply
a weak measurement of the stabilizer $Z$ and a weak rotation around
the X-axis, which depends on the state of the system at that moment.
In the simplified version of that scheme in
Ref.~\cite{Sarovar:2004:052324}, the feedback is condition only on
an estimate of the average measurement current. If at a given moment
the state is somewhere along the $Z$-axis, i.e., $\rho = \alpha
|0\rangle \langle 0| + (1-\alpha) |1\rangle \langle 1|$,
$0\leq\alpha\leq 1$, the effect of a weak measurement would be to
move the state slightly up or down along the axis depending on the
outcome. It is easy to see that the result of such a measurement
does not change the value of $\alpha$ on average, because
$M^Z_+(\epsilon)\rho M^Z_+(\epsilon) +M^Z_-(\epsilon)\rho
M^Z_-(\epsilon) =\rho$. One is then led to ask whether including
feedback could improve the average fidelity. The answer depends on
whether the state lies in the northern or the southern hemisphere of
the Bloch sphere. If the state lies on the $Z$-axis in the northern
hemisphere, it is not possible to improve its fidelity by feedback.
Assuming that the measurement is sufficiently weak so that the
negative outcome $M^Z_-(\epsilon)\rho
M^Z_-(\epsilon)/\textrm{Tr}(M^Z_-(\epsilon)\rho M^Z_-(\epsilon))$ is
still in the northern hemisphere, no unitary operation can bring any
of the two outcomes closer to the north pole since unitary
operations preserve the distance from the center. On the contrary, a
unitary rotation around the X-axis would move both outcomes away
from the $Z$-axis and therefore away from the target.

In the ADL scheme there is no risk for the feedback to decrease the
fidelity with the target state because the feedback is conditioned
on the current state and always tends to increase the fidelity with
the code space; if the state lies on the $Z$-axis in the northern
hemisphere, no rotation would be applied. However, during initial
times that scheme would not be helpful for increasing the average
value of $\alpha$ either, because a weak measurement keeps the state
on the $Z$-axis in the northern hemisphere. If we go to the
continuous limit, $\epsilon\rightarrow 0$, the Wiener parameter is
normally distributed and during an infinitesimal time step the state
may enter the southern hemisphere, but with a negligible
probability. Thus during initial times, the scheme would not be
helpful with respect to the average fidelity, and only after the
probability for the state to enter the southern hemisphere becomes
significant will it start to have an effect. This intuition is
confirmed by the numerical simulations of a generalization of this
protocol to multi-qubit codes presented in
Ref.~\cite{Ahn:2002:042301}.

In the scheme in Ref.~\cite{Sarovar:2004:052324}, the feedback is
not conditioned on the state but on an estimate of the average
measurement current \eqref{meascurr}. The idea is that by filtering
the noisy measurement data obtained during some short time interval
before a given moment $t$, we can try to obtain an estimate of the
average change of $x(t)$ with time at that moment, i.e., an estimate
of $\langle Z\rangle_t$. But clearly such an estimate cannot be
precise, because it would mean that we could measure the expectation
value of an observable almost without disturbing the state.
Therefore, any such estimate inevitably carries imprecision. For
example, it could be that the state of the system is
$|0\rangle\langle 0|$ but we obtain a sequence of negative outcomes
which give rise to the effective measurement operator $M^Z(x) =
\sqrt{I+\tanh(x) Z}$ with $x<0$. This can occur with finite
probability and it would suggest that the state lies in the southern
hemisphere, while the state will remain $|0\rangle\langle 0|$ under
this measurement. In such a case, this scheme would apply a rotation
which would take the state away from the target state, i.e., during
short initial times this scheme could have a negative effect.
Nevertheless, as time progresses, more and more trajectories enter
the southern hemisphere and the scheme may lead to an improvement of
the average fidelity with the target state at later times. Indeed,
numerical simulations have confirmed the efficiency of this scheme
and its generalization to multi-qubit codes in certain parameter
regimes \cite{Sarovar:2004:052324}.

We point out that the two general strategies for the protection of a
qubit that we considered---the one involving continuous measurement
of the $X$ operator and direct feedback (the quantum-jump scheme),
and the one involving continuous measurements of the $Z$ operator
and indirect feedback (the ADL and similar schemes)---strongly
resemble two optimal protocols for the purification of a qubit
discussed in Refs.~\cite{Jacobs:2004:355} and
\cite{Wiseman:2006:90}. In Ref.~\cite{Jacobs:2004:355} it was shown
that the fastest increase on average of the purity of a single qubit
using weak measurements is achieved if the qubit is measured in a
basis perpendicular to the axis in the Bloch sphere that connects
the current state with the center of the sphere. If we assume that
we can apply fast unitary rotations on the time scale of the
measurements, the fastest preparation of a qubit in the state
$|0\rangle\langle 0|$ can be achieved by measuring the state in the
eigenbasis of $X$, and after every weak measurement apply a rotation
around the $Y$-axis that brings this state to the $Z$-axis. This is
almost the same as the quantum-jump scheme, except that we did not
assume that we can apply an arbitrarily strong and precise rotation
that brings each outcome on the $Z$-axis, but only a rotation which
would bring the state to the north pole if it was there before the
measurement.

In Ref.~\cite{Jacobs:2004:355}, on the other hand, it was shown that
if we are interested in the \textit{average time} that it would take
to purify the qubit to a certain degree, we have to measure it along
the axis that connects it with the center of the Bloch sphere.
Again, if we assume that we can apply arbitrarily fast rotations,
the optimal average time for preparing a qubit in the state
$|0\rangle\langle 0|$ with some precision can be achieved if we
measure the qubit in the eigenbasis of $Z$ and whenever the qubit enters the
southern hemisphere, apply rotations around the $X$-axis that bring
it to the northern half of the $Z$-axis. The difference of the ADL
scheme from this approach is again that the ADL scheme does not
assume infinitely fast and precise rotations. Thus we see that the
two competitive error-correction schemes we discussed can be
regarded as originating from two optimal protocols for the
preparation of a qubit in a known state---one that optimizes the
average fidelity with the target state, and another that optimizes
the average time to reach the target state.

Of course, this does not mean that the two schemes we described are
optimal for the resources they use. In the quantum jump scheme, for
example, we discard information about the outcome of the measurement
after every feedback operation. If we keep this information and
estimate the current state, we can in principal improve the
performance of the scheme. Let us say that the state is somewhere
far from the $Z$-axis. Since each of the outcomes of the weak
measurement change the state by a small amount, after either outcome
we will have to apply rotations in the same direction in order to
bring the state closer to the $Z$-axis. If we do not keep track of
the actual state, however, we would apply rotations in opposite
directions after the two different outcomes. But it turns out that
the improvement we can gain by keeping track of the actual states is
small. It can be verified that even if we assume that we are able to
apply infinitely fast and precise rotations, i.e., that we can bring
the state on the $Z$-axis after every weak measurement outcome, if
the measurement strength is fixed, the correction to the quantity
$(1-\alpha_*^{\rm M})$ (Eq.~\eqref{attractor}) we can obtain is of
order $O((1-\alpha_*^{\rm M})^2)$. But as we argued in
Sec.~\ref{sectionENCODEDBASIS} and will discuss further in
Sec.~\ref{exact}, this is the quantity that is responsible for the
effective decrease of the error rate in a general code. In that
sense, the performance of the quantum-jump scheme is very close to
optimal when $(1-\alpha_*^{\rm M})$ is small, even though the scheme
requires no side information processing. Note, however, that we
assumed that at the level of a single weak operation we can ensure a
particular relation between the measurement strength and the
strength of the correcting rotation---Eq.~\eqref{epsilonprime}. If
we cannot apply a sufficiently strong rotation to keep the state
$|0\rangle\langle 0|$ invariant, the equilibrium fidelity with the
target state $\alpha_*$ would be lower.

\subsection{Generalizations to multi-qubit codes}

A natural extension of the single-qubit schemes with indirect
feedback to non-trivial codes can be obtained simply by applying
these schemes to the syndrome qubits in the encoded basis with the
purpose of keeping each of them close to the state $|0\rangle\langle
0|$. It is not hard to see that the operators $Z_i^s$ on the gauge
qubits in the encoded basis are actually the stabilizer generators
for the code. For example, by applying the inverse of the basis
transformation for the bit-flip code described in
Sec.~\ref{sectionENCODEDBASIS}, one can see that the operators
$I^A\otimes Z^s_1\otimes I^s_2$ and $I^A\otimes I^s_1\otimes Z^s_2$
correspond to the generators $ZZI$ and $ZIZ$, respectively.

The Hamiltonians $X_i^s$ needed for the feedback, however, do not
have simple forms in the original basis. In particular, for the
bit-flip code, the operators $I^A\otimes X^s_1\otimes I^s_2$ and
$I^A\otimes I^s_1\otimes X^s_2$ correspond to
$\frac{1}{2}XIX+\frac{1}{2}YIY+\frac{1}{2}ZXZ+\frac{1}{2}IXI$ and
$\frac{1}{2}XXI+\frac{1}{2}YYI+\frac{1}{2}ZZX+\frac{1}{2}IIX$,
respectively. The models considered in Refs.~\cite{Ahn:2002:042301,
Sarovar:2004:052324, Chase:2008:032304} also measure continuously
the stabilizer generators of the code, but the feedback Hamiltonians
are assumed to be single-qubit operators in the original basis.
However, note that in the general formulation of the ADL
scheme---Eq.~\eqref{ADL}---the correcting Hamiltonians $H_r$ are not
specified, and in that sense the possibility we discuss here can be
regarded as a special case of the ADL scheme.

In the case of the bit-flip code, the authors in
Refs.~\cite{Ahn:2002:042301, Sarovar:2004:052324} take the
correcting Hamiltonians to be $XII$, $IXI$ and $IIX$. This choice is
motivated one hand by its analogy with the strong version of the
error-correcting operation for this code, and on the other by its
simplicity. In the encoded basis, however, these operators are
correlated and act on subsystem $\mathcal{H}^A$ as well. More
precisely, $XII$, $IXI$ and $IIX$ are equal to
$\frac{1}{2}IXX-\frac{1}{2}IYY+\frac{1}{2}XXX+\frac{1}{2}XYY$,
$\frac{1}{2}IXZ+\frac{1}{2}IXI+\frac{1}{2}XXI-\frac{1}{2}XXZ$, and
$\frac{1}{2}IZX+\frac{1}{2}IIX+\frac{1}{2}XIX-\frac{1}{2}XZX$,
respectively. Naturally, since the code is designed to correct
single-qubit bit flips, these operators leave the factor
$\mathcal{H}^A$ in the code space $\mathcal{H}^A\otimes
|0\rangle^s_1\otimes |0\rangle^s_2$ invariant by definition. A
similar property holds for codes that can correct arbitrary
single-qubit errors. But these operators can introduce errors to the
code subsystem through their non-trivial action on the orthogonal
complement of the code space. In particular, imagine that the system
undergoes just a single perfectly correctable error, say, a single
bit flip. Then a strong error correcting operation must be able to
correct it. But if we apply a continuous scheme in which the
correcting Hamiltonians act non-trivially on the complement of the
code space, this scheme would generally apply non-trivial
transformations to the subsystem $\mathcal{H}^A$ in the error
subspace, which are by definition uncorrectable. (Note that this
cannot occur with a scheme which uses operations acting locally on
the syndrome qubits.) Nevertheless, in the case of continuous
decoherence where uncorrectable errors inevitably arise, this
property is not of crucial significance. As we argued earlier, the
way CTQEC works is by keeping the weight outside the code space
small, which suppresses the effective accumulation of uncorrectable
errors. As long as the scheme is able to keep that weight small, it
will still have an effect according to our earlier arguments.
Indeed, numerical simulations show that with the use of single-qubit
feedback Hamiltonians one can achieve a significant improvement of
the codeword fidelity with respect to that of an unprotected qubit
and outperform the approach of single-shot error correction in
various regimes. For details about the numerical results, we refer
the reader to Refs.~\cite{Ahn:2002:042301, Sarovar:2004:052324,
Chase:2008:032304}.

\section{Quantum jumps for Markovian and non-Markovian noise}\label{exact}

In this section we will look at the performance of the quantum-jump
scheme in the cases of Markovian and non-Markovian decoherence. We
will consider the bit-flip code in the case of simple noise models
for which the evolution is exactly solvable. The conclusions we
obtain, however, hold for general codes and noise models.

\subsection{Markovian decoherence}

The model described by Eq.~\eqref{errorcorrectionequation}
represents the noise as driven by a Lindblad generator, which is
valid under the Markovian assumption of bath correlation times that
are much shorter than any characteristic time scale of the system
\cite{Breuer:2002:Oxford}. In the case of protecting a single qubit
from Markovian bit-flip decoherence, we already found the solution
for this model---Eq.~\eqref{MSQS}. We saw that the equilibrium
fidelity to which the qubit decays scales as $1/\kappa$ for large
error-correction rates $\kappa$.

For the bit-flip code, we will assume that all qubits decohere
through identical independent bit-flip channels, i.e.,
$\mathcal{L}(\rho)$ is of the form \eqref{Lbitflip} with
$\lambda_1=\lambda_2=\lambda_3=\lambda$. Then one can verify that
the density matrix at any moment can be written
\begin{equation}
\rho(t) = a(t)\rho(0)+b(t)\rho_{1}+c(t)\rho_{2}+d(t)\rho_{3},
\label{rhooft}
\end{equation}
where
\begin{eqnarray}
\rho_{1}&=&\frac{1}{3}(X_1\rho(0)X_1+X_2\rho(0)X_2 +
X_3\rho(0)X_3),\notag\\
\rho_{2}&=&\frac{1}{3}(X_1X_2\rho(0)X_1X_2+
X_2X_3\rho(0)X_2X_3+X_1X_3\rho(0)X_1X_3),\notag\\
\rho_{3} &=& X_1X_2X_3\rho(0) X_1X_2X_3,
\end{eqnarray}
are equally-weighted mixtures of single-qubit, two-qubit and
three-qubit errors on the original state.

The evolution of the system subject to decoherence plus error
correction is described by the following system of first-order
linear differential equations:
\begin{eqnarray}
\frac{da(t)}{dt}&=&-3\lambda a(t) + (\lambda+\kappa)b(t),\notag\\
\frac{db(t)}{dt}&=&3\lambda a(t)- (3\lambda+\kappa)b(t) + 2
\lambda c(t),\notag\\
\frac{dc(t)}{dt}&=&2\lambda b(t)-
(3\lambda+\kappa)c(t)+3\lambda d(t),\notag\\
\frac{dd(t)}{dt}&=&(\lambda+\kappa)c(t)-3\lambda
d(t).\label{equations}
\end{eqnarray}
The exact solution was found in \cite{Paz:1998:355} and we will not
present it here. We only note that for the initial conditions
$a(0)=1, b(0)=c(0)=d(0)=0$, the exact solution for the weight
outside the code space is
\begin{equation}
b(t)+c(t)=\frac{3}{4+r}(1-e^{-(4+r)\gamma t}),
\end{equation}
where $r=\kappa/\lambda$. We see that similarly to what we obtained
for the single-qubit code, the weight outside the code space quickly
decays to its asymptotic value $\frac{3}{4+r}$ which scales as
$1/r$. But note that this value is roughly three times greater than
that for the single-qubit model. This corresponds to the fact that
there are three single-qubit channels. More precisely, it can be
verified that if for a given $\kappa$ the uncorrected weight by the
single-qubit scheme is small, then the uncorrected weight by a
multi-qubit code using the same $\kappa$ and the same kind of
decoherence for each qubit, scales approximately linearly with the
number of qubits. Similarly, the ratio $r$ required to preserve a
given overlap with the code space scales linearly with the number of
qubits in the code.

The most important difference from the single-qubit model is that
in this model there are non-correctable errors that cause a decay
of the state inside the code space. Due to the finiteness of the
resources employed by our scheme, there always remains a finite
portion of the state outside the code space, which gives rise to
non-correctable three-qubit errors. To understand how the state
decays inside the code space, one can ignore terms of the order of
the weight outside the code space in the exact solution. The result is
\begin{gather}
a(t)\approx  \frac{1+e^{-\frac{6}{r}2\gamma t}}{2}, \hspace{0.3cm}
b(t)\approx 0,\hspace{0.3cm} c(t) \approx 0, \hspace{0.3cm}
d(t)\approx \frac{1-e^{-\frac{6}{r}2\gamma t}}{2}.
\end{gather}
Comparing with the expression for the fidelity of a single
decaying qubit without error correction which can be seen from
\eqref{MSQS} for $\kappa=0$, we see that the encoded qubit decays
roughly as if subject to bit-flip decoherence with rate
$\frac{6}{r}\gamma$. Therefore, for large $r$ this
error-correction scheme can reduce the rate of decoherence
approximately $\frac{r}{6}$ times. In the limit $r \rightarrow
\infty$, it leads to perfect protection of the state for all
times.

\subsection{Non-Markovian decoherence}\label{nM}

\subsubsection{The Zeno effect. Error correction versus error prevention}

The effect of freezing of the evolution in the limit of infinite
error-correction rate bears a strong similarity to the quantum Zeno
effect \cite{Mishra:1977:756}, where frequent measurements slow down
the evolution of a system and freeze the state in the limit where
they are applied continuously. The Zeno effect arises when the
system and its environment are initially decoupled and they undergo
a Hamiltonian-driven evolution, which leads to a quadratic change
with time of the state during the initial moments
\cite{Nakazato:1996:247} (the so-called Zeno regime). Let the
initial state of the system plus the bath be $\rho^{SB}(0)=|0\rangle
\langle 0|^S\otimes\rho^B(0)$. For small times, the fidelity
$\alpha=\textrm{Tr}\{(|0\rangle\langle0|^S\otimes
I^B)\rho^{SB}(t)\}$ of the system's density matrix with the initial
state can be approximated as
\begin{equation}
\alpha(t)= 1-C t^2+\mathcal{O}(t^3).\label{Zeno}
\end{equation}
In terms of the Hamiltonian $H^{SB}$ acting on the entire system,
the coefficient $C$ is
\begin{eqnarray}
C&=&\textrm{Tr}\{(H^{SB})^2|0\rangle\langle0|^S\otimes
\rho^B(0)\}\notag\\
&&-\textrm{Tr}\{H^{SB}|0\rangle\langle0|^S\otimes I^B H^{SB}
|0\rangle\langle0|^S\otimes \rho^B(0)\}.\label{C}
\end{eqnarray}
According to \eqref{Zeno}, if after a time step $\Delta t$ the
state is measured in an orthogonal basis which involves the
initial state, the probability for not projecting it on the
initial state is of order $\mathcal{O}(\Delta t^2)$. Thus if the
state is continuously measured ($\Delta t \rightarrow 0$), this
prevents the system from evolving.

It has been proposed to utilize the quantum Zeno effect in schemes
for error prevention \cite{Zurek:1984:391, Barenco:1997:1541,
Vaidman:1996:1745}, in which an unknown encoded state is protected
from errors simply by frequent measurements that keep it inside the
code space. From the point of view of the encoded basis, this
approach can be understood as measuring the operators $Z^s_i$ which
prevents the syndrome qubits from leaving the state
$|00...0\rangle$. The approach is similar to error correction, in
that the errors for which the code is designed send a codeword to a
space orthogonal to the code space. The difference is that the
subsystem containing the protected information generally does not
remain invariant under the errors, since the procedure does not
involve correction of errors but only their prevention. In
\cite{Vaidman:1996:1745} it was shown that with this approach it is
possible to use codes of smaller redundancy than those needed for
error correction and a four-qubit encoding of a qubit was proposed,
which is capable of preventing arbitrary independent errors arising
from Hamiltonian interactions. The workings of this approach are
based on the existence of a Zeno regime, and fail if we assume
Markovian decoherence for all times. This is because the errors
emerging during a time step $dt$ in a Markovian model are
proportional to $dt$ and they accumulate with time if not corrected.

By the above observations, error correction can achieve results in
noise regimes where error prevention fails. Of course, this
advantage is at the expense of a more complicated procedure---in
addition to the measurements that constitute error prevention, error
correction involves correcting unitaries, and in general is based on
codes with higher redundancy. At the same time, we see that in the
Zeno regime it is possible to reduce decoherence using weaker
resources than those needed for Markovian noise. This suggests that
in this regime error correction may exhibit higher performance than
it does for Markovian decoherence. In many situations of practical
significance, the memory of the environment cannot be neglected, and
the evolution is highly non-Markovian
\cite{Breuer:2002:Oxford,Quang:1997:5238, Breuer:2004:045323,
Krovi:2007:052117}. Furthermore, no evolution is strictly Markovian
and for a system initially decoupled from its environment a Zeno
regime is always present, short though it may be
\cite{Nakazato:1996:247}. Therefore, if the time resolution of
error-correcting operations is high enough so that they ``see'' the
Zeno regime, this could give rise to different behavior.

One important difference between Markovian and non-Markovian noise
is that, in the latter case, the error correction and the effective
noise on the reduced density matrix of the system cannot be treated
as independent processes. One could derive an equation for the
effective evolution of the system alone subject to interaction with
the environment, such as the Nakajima-Zwanzig
\cite{Nakajima:1958:948, Zwanzig:1960:1338} or the
time-convolutionless (TCL) \cite{Shibata:1977:171, Shibata:1980:891}
master equations, but the generator of transformations at a given
moment in general will depend (implicitly or explicitly) on the
entire history up to this moment. Therefore, adding error correction
can affect the effective error model nontrivially. This means that
in order to describe the evolution of a system subject to
non-Markovian decoherence and error correction, one either has to
derive an equation for the effective evolution taking into account
error correction from the very beginning, or one has to look at the
evolution of the entire system including the bath, where the error
generator and the generator of error correction can be considered
independent. In the latter case, for sufficiently small $\tau_c$,
the evolution of the entire system including the bath can be
described by
\begin{equation}
\frac{d \rho}{dt}=-i[H, \rho(t)]+\kappa
\mathcal{J}(\rho),\label{NMerrorcorrectionequation}
\end{equation}
where $\rho$ is the density matrix of the system plus the bath, $H$
is the total Hamiltonian, and the error-correction generator
$\mathcal{J}$ acts locally on the encoded system. We will consider a
description in terms of Eq.~\eqref{NMerrorcorrectionequation} for a
sufficiently simple bath model which allows us to find a solution
for the evolution of the entire system. To gain understanding of how
the scheme works, we will again look at the single-qubit model
first.

\subsubsection{The single-qubit code}\label{sectionSQNM}

We choose the simple scenario of a system coupled to a single bath
qubit via the Hamiltonian
\begin{equation}
H=\gamma X\otimes X,
\end{equation}
where $\gamma$ is the coupling strength. This can be a good
approximation for situations in which the coupling to a single
spin from the bath dominates over other interactions
\cite{Krovi:2007:052117}.

\begin{figure}[h]
\begin{center}
\includegraphics[width=4in]{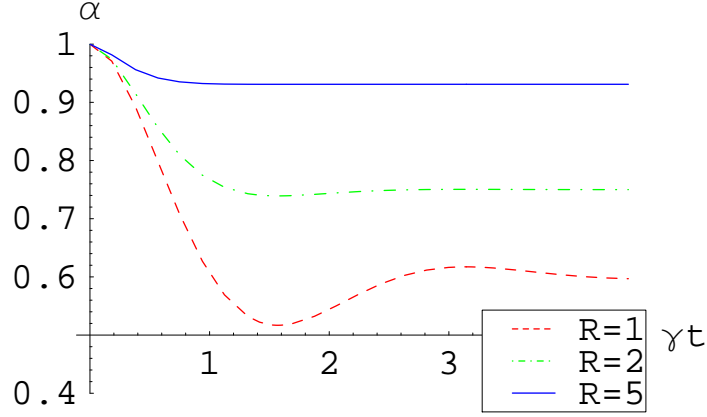}
\caption{Fidelity of the single-qubit code with continuous
bit-flip errors and correction, as a function of dimensionless
time $\gamma t$, for three different values of the ratio
$R=\kappa/\gamma$.} \label{fig1}
\end{center}
\end{figure}

We assume that the bath qubit is initially in the maximally mixed
state, which can be thought of as an equilibrium state at high
temperature. From \eqref{NMerrorcorrectionequation} one can verify
that if the system is initially in the state $|0\rangle$, the state
of the system plus the bath at any moment will have the form
\begin{eqnarray}
\rho(t) = \left(\alpha (t) |0\rangle \langle 0|
+(1-\alpha(t))|1\rangle \langle 1|\right)\otimes \frac{I}{2}-
\beta(t)Y \otimes \frac{X}{2}.
\end{eqnarray}
In the tensor product, the first operator belongs to the Hilbert
space of the system and the second to the Hilbert space of the bath.
We have $\alpha(t) \in [0,1]$, and
$|\beta(t)|\le\sqrt{\alpha(t)(1-\alpha(t))}, \beta(t)\in R$. The
reduced density matrix of the system has the same form as the one
for the Markovian case. The part proportional to $\beta(t)$ can be
thought of as a ``hidden'' part, which nevertheless plays an
important role in the error-creation process, since errors can be
thought of as being transferred to the ``visible'' part from the
``hidden'' part (and vice versa). This can be seen from the fact
that during an infinitesimal time step $dt$, the Hamiltonian changes
the parameters $\alpha$ and $\beta$ as follows:
\begin{gather}
\alpha\rightarrow \alpha-2\beta \gamma dt ,\notag\\
\beta \rightarrow \beta +(2\alpha-1)\gamma dt .\label{sqe}
\end{gather}
The effect of an infinitesimal error-correcting operation is
\begin{gather}
\alpha \rightarrow \alpha + (1-\alpha)\kappa dt,\notag\\
\beta\rightarrow \beta-\beta\kappa dt.
\end{gather}
Note that the ``hidden'' part is also being acted upon. Putting it
all together, we get the system of equations
\begin{gather}
\frac{d \alpha(t)}{dt}=\kappa(1-\alpha(t))-2\gamma \beta(t),\notag\\
\frac{d
\beta(t)}{dt}=\gamma(2\alpha-1)-\kappa\beta(t)\label{equation2}.
\end{gather}
The solution for the fidelity $\alpha(t)$ is
\begin{gather}
\alpha(t)=\frac{2\gamma^2 + \kappa^2}{4\gamma^2+\kappa^2}
+e^{-\kappa
t}\left(\frac{\kappa\gamma}{4\gamma^2+\kappa^2}\sin{2\gamma
t}+\frac{2\gamma^2}{4\gamma^2+\kappa^2}\cos{2\gamma
t}\right).\label{singlequbitsolution}
\end{gather}
We see that as time increases, the fidelity stabilizes at the
value
\begin{equation}
\alpha_*^{NM}= \frac{2+R^2}{4+R^2}=1-\frac{2}{4+R^2},
\end{equation}
where $R=\kappa/\gamma$ is the ratio between the error-correction
rate and the coupling strength. Fig.~\ref{fig1} shows the
fidelity as a function of the dimensionless parameter $\gamma t$
for three different values of $R$. For error-correction rates
comparable to the coupling strength ($R=1$), the fidelity
undergoes a few partial recurrences before it stabilizes close to
$\alpha_*^{NM}$. For larger $R=2$, however, the oscillations are
already heavily damped and for $R=5$ the fidelity seems confined
above $\alpha_*^{NM}$. As $R$ increases, the evolution becomes
closer to a decay like the one in the Markovian case.

A remarkable difference, however, is that the asymptotic weight
outside the code space ($1-\alpha_*^{NM}$) decreases with $\kappa$
as $1/\kappa^2$, whereas in the Markovian case the same quantity
decreases as $1/\kappa$. The asymptotic value can be obtained as an
equilibrium point at which the infinitesimal weight flowing out of
the code space during a time step $dt$ is equal to the weight
flowing into it. The latter corresponds to vanishing right-hand
sides in equations \eqref{equation1} and \eqref{equation2}. In Sec.~\ref{sectionROLEOFZENO} we will
see that the difference in that quantity for the two different types
of decoherence arises from the difference in the corresponding
evolutions during initial times.

\subsubsection{The three-qubit bit-flip code}\label{sectionMQNM}

We will consider a model where each qubit independently undergoes the
same kind of non-Markovian decoherence as the one we studied for
the single-qubit code. Here the system we look at consists of six
qubits---three for the codeword and three for the environment. We
assume that all system qubits are coupled to their corresponding
environment qubits with the same coupling strength, i.e., the
Hamiltonian is
\begin{equation}
H=\gamma\overset{3}{\underset{i=1}{\sum}}X^S_i\otimes
X^B_i,\label{Hamiltonian}
\end{equation}
where the operators $X^S$ act on the system qubits and $X^B$ act on
the corresponding bath qubits which are initially in the maximally
mixed state. The subscripts label on which particular qubit they
act. Obviously, the types of effective single-qubit errors on the
density matrix of the system that can result from this Hamiltonian
at any time, CP or not, will have operator elements which are linear
combinations of the identity and $X^S$. According to the
error-correction conditions for non-CP maps obtained in
Ref.~\cite{Shabani:07081953}, these errors are correctable by the
code. Considering the form of the Hamiltonian \eqref{Hamiltonian}
and the error-correcting map, one can see that the density matrix of
the entire system at any moment is a linear combination of terms of
the type
\begin{equation}
\varrho_{lmn,pqr}\equiv X_1^lX_2^mX_3^n\rho(0)
X_1^pX_2^qX_3^r\otimes \frac{X_1^{l+p}}{2}\otimes
\frac{X_2^{m+q}}{2} \otimes \frac{X_3^{n+r}}{2}.
\end{equation}
Here the first term in the tensor product refers to the Hilbert
space of the system, and the following three refer to the Hilbert
spaces of the bath qubits that couple to the first, the second and
the third qubits from the code respectively. The power indices
$l,m,n,p,q,r$ take values $0$ and $1$ in all possible
combinations, and $X^1=X$, $X^0=X^2=I$. (Note that
$\varrho_{lmn,pqr}$ should not be mistaken with the components of
the density matrix in the computational basis.) More precisely, we
can write the density matrix in the form
\begin{eqnarray}
\rho(t)&=&\underset{l,m,n,p,q,r}{\sum}(-i)^{l+m+n}(i)^{p+q+r}C_{lmn,pqr}(t)\times
\varrho_{lmn,pqr},\label{fullDM}
\end{eqnarray}
where the coefficients $C_{lmn,pqr}(t)$ are real. The coefficient
$C_{000,000}$ is less than or equal to the codeword fidelity (with
equality when $\rho(0)=|\bar{0}\rangle\langle \bar{0}|$ or
$\rho(0)=|\bar{1}\rangle\langle \bar{1}|$). Since the scheme aims at
protecting an unknown codeword, we will be interested in its
performance in the worst case and we will assume that the codeword
fidelity is $C_{000,000}$.

The exact equations for the coefficients $C_{lmn,pqr}(t)$ and their solutions were obtained in Ref.~\cite{Oreshkov:2007:022318}. Here we will present an approximation which can be obtained by perturbation theory for $\gamma \delta t \ll 1 \ll \kappa \delta t$ \cite{Oreshkov:2007:022318}. The approximate system of equation reads
\begin{gather}
\frac{d C_{000,000}}{dt}=\frac{24}{R^2}\gamma C_{111,000},\notag\\
\frac{d C_{111,000}}{dt}=-\frac{12}{R^2}\gamma
(2C_{000,000}-1).\label{approxeqn}
\end{gather}
Comparing with \eqref{sqe}, we see that the encoded qubit
undergoes approximately the same type of evolution as that of a
single qubit without error correction, but the coupling constant
is effectively decreased $R^2/12$ times. The solution of
\eqref{approxeqn} yields for the codeword fidelity
\begin{equation}
C_{000,000}(t)=\frac{1+\cos (\frac{24}{R^2}\gamma t)}{2}
\label{firstapproxsoln}.
\end{equation}
This solution is valid only with precision
$\mathcal{O}(\frac{1}{R})$ for times $\gamma t \ll R^3$. If one
carries out the perturbation to fourth order in $\gamma$, one
obtains the approximate equations
\begin{gather}
\frac{d C_{000,000}}{dt}=\frac{24}{R^2}\gamma C_{111,000}-\frac{72}{R^3}\gamma (2 C_{000,000}-1),\notag\\
\frac{d C_{111,000}}{dt}=-\frac{12}{R^2}\gamma (2C_{000,000}-1)
-\frac{144}{R^3}\gamma C_{111,000},\label{approxeqn2}
\end{gather}
which yield for the fidelity
\begin{equation}
C_{000,000}(t)=\frac{1+e^{-\frac{144}{R^3}\gamma t}\cos
(\frac{24}{R^2}\gamma t)}{2}.
\end{equation}
We see that in addition to the effective error process which is of
the same type as that of a single qubit, there is an extra Markovian
bit-flip process with rate $\frac{72}{R^3}\gamma$. This Markovian
behavior is due to the Markovian character of our error-correcting
procedure which, at this level of approximation, is responsible for
the direct transfer of weight between $\varrho_{000,000}$ and
$\varrho_{111,111}$, and between $\varrho_{111,000}$ and
$\varrho_{000,111}$. The exponential factor explicitly reveals the
range of applicability of solution \eqref{firstapproxsoln}---with
precision $\mathcal{O}(\frac{1}{R})$, it is valid  only for times
$\gamma t$ of up to order $R^2$. For times of the order of $R^3$,
the decay becomes significant and cannot be neglected. The
exponential factor may also play an important role for short times
of up to order $R$, where its contribution is bigger than that of
the cosine. But in the latter regime the difference between the
cosine and the exponent is of order $\mathcal{O}(\frac{1}{R^2})$,
which is negligible for the precision that we consider.

\begin{figure}[h]
\begin{center}
\includegraphics[width=4in]{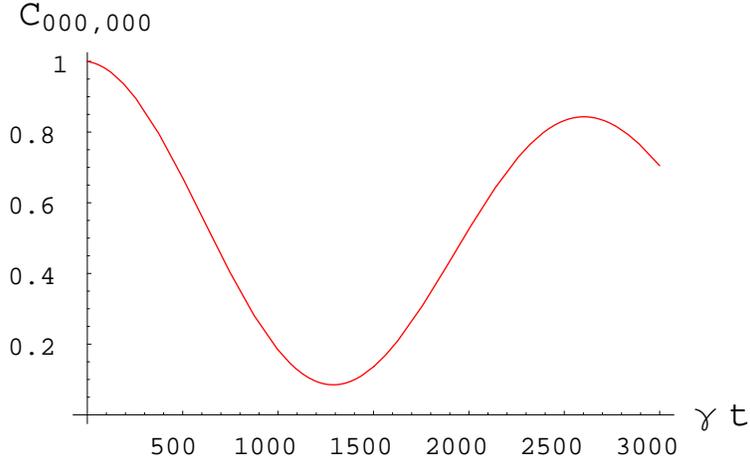}
\caption{Long-time behavior of three-qubit system with bit-flip
noise and continuous error correction.  The ratio of correction
rate to decoherence rate is $R=\kappa/\gamma=100$.} \label{fig3}
\end{center}
\end{figure}

\begin{figure}[h]
\begin{center}
\includegraphics[width=4in]{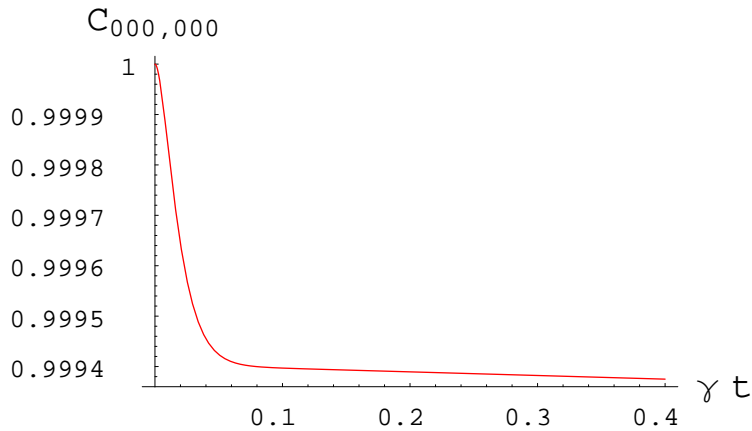}
\caption{Short-time behavior of three-qubit system with bit-flip
noise and continuous error correction.  The ratio of correction
rate to decoherence rate is $R=\kappa/\gamma=100$.} \label{fig4}
\end{center}
\end{figure}

Fig.~\ref{fig3} presents the exact solution for the codeword
fidelity $C_{000,000}(t)$ as a function of the dimensionless
parameter $\gamma t$ for $R=100$. For very short times after the
beginning ($\gamma t \sim 0.1$), one can see a fast but small in
magnitude decay (Fig.\ref{fig4}). The maximum magnitude of this
quickly decaying term obviously decreases with $R$, since in the
limit of $R\rightarrow \infty$ the fidelity should remain constantly
equal to $1$.

We see that in the limit $R\rightarrow \infty$, the evolution
approaches an oscillation with an angular frequency
$\frac{24}{R^2}\gamma$. This is the same type of evolution as that
of a single qubit interacting with its environment, but the coupling
constant is effectively reduced $R^2/12$ times. While the coupling
constant can serve to characterize the decoherence process in this
particular case, such a description is not valid in general. As a
general measure of the effect of noise one can use the
instantaneous rate of decrease of the codeword fidelity $F_{cw}$ (in
our case $F_{cw}=C_{000,000}$):
\begin{equation}
\Lambda(F_{cw}(t)) = -\frac{dF_{cw}(t)}{dt}.\label{errorrate}
\end{equation}
This quantity does not coincide with the decoherence rate in the
Markovian case (which can be defined naturally from the Lindblad
equation), but it is a good estimate of the rate of loss of fidelity
and can be used for any decoherence model. We will refer to it
simply as an error rate. Since the goal of error correction is to
preserve the codeword fidelity, the quantity \eqref{errorrate} is a
useful indicator of the performance of a given scheme. Note that
$\Lambda (F_{cw})$ is a function of the codeword fidelity and
therefore it makes sense to use it for a comparison between
different cases only for identical values of $F_{cw}$. For our
example, the fact that the coupling constant is effectively reduced
approximately $R^2/12$ times implies that the error rate for a given
value of $F_{cw}$ is also reduced $R^2/12$ times. Similarly, the
reduction of $\lambda$ by the factor $r/6$ in the Markovian case
implies a reduction of $\Lambda$ by the same factor. We see that the
effective reduction of the error rate increases quadratically as
$\kappa^2$ in the non-Markovian case, whereas it increases only
linearly as $\kappa$ in the Markovian case.

\subsection{The role of the Zeno regime}\label{sectionROLEOFZENO}

The effective continuous evolution \eqref{approxeqn} is derived
under the assumption $\gamma \delta t \ll 1 \ll \kappa \delta t $.
The first inequality implies that $\delta t$ can be considered
within the Zeno time scale of the system's evolution without error
correction. On the other hand, from the relation between $\kappa$
and $\tau_c$ in \eqref{tauc} we see that $\tau_c\ll\delta t$.
Therefore, the time for implementing a weak error-correcting
operation has to be sufficiently small so that on the Zeno time
scale the error-correction procedure can be described approximately
as a continuous Markovian process. This suggests a way of
understanding the quadratic enhancement in the non-Markovian case
based on the properties of the Zeno regime.

Let us consider again the single-qubit code from
Sec.~\ref{sectionSQNM}, but this time let the error model be any
Hamiltonian-driven process. We assume that the qubit is initially in
the state $|0\rangle\langle 0|$, i.e., the state of the system
including the bath has the form $\rho(0)=|0\rangle \langle
0|\otimes\rho_B(0)$. For times smaller than the Zeno time $\delta
t_Z$, the evolution of the fidelity without error correction can be
described by \eqref{Zeno}. Equation \eqref{Zeno} naturally defines
the Zeno regime in terms of $\alpha$ itself:
\begin{equation}
\alpha\geq \alpha_Z \equiv 1-C\delta t_Z^2.
\end{equation}
For a single time step $\Delta t \ll \delta t_Z$, the change in
the fidelity is
\begin{equation}
\alpha\rightarrow \alpha-2\sqrt{C}\sqrt{1-\alpha}\Delta
t+\mathcal{O}(\Delta t^2).\label{singlestepdecoh}
\end{equation}
On the other hand, the effect of error correction during time
$\Delta t$ is
\begin{equation}
\alpha \rightarrow \alpha+\kappa (1-\alpha)\Delta t
+\mathcal{O}(\Delta t^2),\label{singlestepcorr}
\end{equation}
i.e., it tends to oppose the effect of decoherence.  If both
processes happen simultaneously, the effect of decoherence will
still be of the form \eqref{singlestepdecoh}, but the coefficient
$C$ may vary with time. This is because the presence of
error-correction opposes the decrease of the fidelity, and
consequently can lead to an increase in the time for which the
fidelity remains within the Zeno range. If this time is sufficiently
long, the state of the environment could change significantly under
the action of the Hamiltonian, thus giving rise to a different value
for $C$ in \eqref{singlestepdecoh} according to \eqref{C}. Note that
the strength of the Hamiltonian puts a limit on $C$, and therefore
this constant can vary only within a certain range. The equilibrium
fidelity $\alpha_*^{NM}$ that we obtained for the error model in
Sec.~\ref{sectionSQNM} can be thought of as the point at which the
effects of error and error correction cancel out. For a general
model, where the coefficient $C$ may vary with time, this leads to a
quasi-stationary equilibrium. From \eqref{singlestepdecoh} and
\eqref{singlestepcorr}, one obtains the equilibrium fidelity
\begin{equation}
\alpha_*^{NM}\approx 1-\frac{4C}{\kappa^2}.
\end{equation}
In agreement to the result in Sec.~\ref{sectionSQNM}, the
equilibrium fidelity differs from $1$ by a quantity proportional to
$1/\kappa^2$. If one assumes a Markovian error model, for short
times the fidelity changes linearly with time which leads to
$1-\alpha_*^{M}\propto 1/\kappa$. Thus the difference can be
attributed to the existence of a Zeno regime in the non-Markovian
case.

This argument readily generalizes to the case of non-trivial codes
if we look at the picture in the encoded basis. There each syndrome
qubit undergoes a Zeno-type evolution and so do the abstract qubits
associated with each error syndrome. Then using only the properties
of the Zeno behavior as we did above, we can conclude that the
weight outside the code space will be kept at a quasi-stationary
value of order $1/\kappa^2$. As we argued in
Sec.~\ref{sectionENCODEDBASIS}, this in turn would lead to an
effective decrease of the uncorrectable error rate at least by a
factor proportional to $1/\kappa^2$.

Finally, let us make a remark about the resources needed to achieve
the effect of quadratic reduction of the error rate. As it was
pointed out, there are two conditions involved---one concerns the
magnitude of the error-correction rate, the other concerns the time
resolution of the weak error-correcting operations. Both of these
quantities should be sufficiently large. There is, however, an
interplay between the two, which involves the strength of the
interaction required to implement the weak error correcting map
\eqref{wm}. Let us imagine that the weak map is implemented by making
the system interact weakly with an ancilla in a given state, after
which the ancilla is discarded. The error correction procedure
consists of a sequence of such interactions and can be thought of as
a cooling process. If the time for which a
single ancilla interacts with the system is $\tau_c$, one can verify
that the parameter $\varepsilon$ in \eqref{wm} would be proportional
to $g^2\tau_c^2$, where $g$ is the coupling strength between the
system and the ancilla. From \eqref{tauc} we then obtain that
\begin{equation}
\kappa \propto g^2\tau_c.
\end{equation}
The two parameters that can be controlled are the interaction time
and the interaction strength, and they determine the
error-correction rate. Thus, if $g$ is kept constant, a decrease in
the interaction time $\tau_c$ leads to a proportional decrease in
$\kappa$ which may be undesirable. Therefore, in order to achieve a
good working regime, one generally may need to adjust both $\tau_c$
and $g$. But in some situations decreasing $\tau_c$ alone can prove
advantageous, since this may lead to a time resolution that reveals
the non-Markovian character of an error model that was previously
treated as Markovian. Then the quadratic enhancement of the
performance as a function of $\kappa$ may compensate the decrease in
$\kappa$, thus leading to a seemingly paradoxical result---better
performance with a lower error-correction rate.

\section{Outlook}\label{discussion}

In this chapter we saw that the subsystem principle can be useful
for understanding various aspects of the workings of CTQEC and its
performance under different noise models, as well as for the design
of CTQEC protocols using protocols for the protection of a known
state. However, further research is needed to understand how to
construct optimal CTQEC protocols. In the case of the quantum-jump
model, the code-space fidelity reaches a quasi-equilibrium value
which can be used to estimate the performance of the scheme. It
would be interesting to see whether an analogue of the equilibrium
fidelity exists for schemes with indirect feedback. This could prove
useful since stochastic evolutions are generally too complicated for
analytical treatment. The equilibrium code-space fidelity can be
useful also in assessing the performance of CTQEC under
non-Markovian decoherence, where the description of the evolution of
a system subject to CTQEC can be difficult due to the large number
of environment degrees of freedom.

We discussed two main methods for obtaining CTQEC protocols from
protocols for the protection of a single qubit: one based on the
application of single-qubit protocols to the separate syndrome
qubits, and another based on the application of single-qubit
protocols to qubit subspaces associated with the different
syndromes. An interesting question is whether the performance of
CTQEC protocols obtained by these methods can be related to the
performance of the underlying single-qubit protocols. A difficulty
in the case of indirect feedback is that the noise in the encoded
basis is correlated, and the effective noise on a given qubit can
depend on the outcomes of the measurements on the rest of the
qubits.

Another interesting direction for future investigation is to explore
CTQEC for specific physical models and limitations of the control
parameters (for a recent work, see Ref.~\cite{Kerckhoff:0812:1246}).
We saw that applying single-qubit schemes to the syndrome qubits in
the encoded basis generally requires multi-qubit operations in the
original basis, but numerical simulations show that single-qubit
feedback Hamiltonians in the original basis are also efficient. It
would be interesting to see whether it is possible to construct
efficient CTQEC protocols for non-trivial codes assuming only one-
and two-qubit Hamiltonians. The ability to apply CTQEC with
Hamiltonians of limited locality would be important for the
scalability of this approach.

So far, CTQEC has been considered only as a method of protecting
quantum memory. A natural next step is to combine this approach with
universal quantum computation. An important question in this respect
is whether CTQEC can be made fault tolerant. In the theory of
quantum fault tolerance, logical operations and error correction are
implemented mainly in terms of transversal operations between
physical qubits from different blocks, where the basic operations
are assumed to be discrete. Is something similar possible for CTQEC?
One way of approaching this problem could be to look for
fault-tolerant implementations of a universal set of weak operations
using only weak transversal unitary operations and projective
ancilla measurements.

Undoubtedly, the area of CTQEC offers a variety of interesting
problems for future investigation. As quantum operations with
limited strength or limited rate are likely to be the tools
available in many quantum computing architectures in the near term,
developing further the approach to protecting quantum information
from noise via continuous-time feedback seems a promising direction
for research.

\acknowledgements{O.O. acknowledges the support of the European Commission under the Marie Curie Intra-European Fellowship Programme (PIEF-GA-2010-273119).  This research was supported in part by the Spanish
MICINN (Consolider-Ingenio QOIT). }

\bibliographystyle{plain}
\bibliography{myrefs}

\begin{thebibliography}{10}

\bibitem{Barenco:1997:1541}
{A. Barenco, A. Berthiaume, D. Deutsch, A. Eckert, R. Jozsa and C.
  Macchiavello}.
\newblock {Stabilization of quantum computations by symmetrization}.
\newblock {\em {SIAM Journal on Computing}}, 26:1541, 1997.

\bibitem{Leggett:1989:2325}
{A. J. Leggett}.
\newblock {Comment on ``How the result of a measurement of a component of the
  spin of a spin-(1/2 particle can turn out to be 100''}.
\newblock {\em {Phys. Rev. Lett.}}, 62:2325, 1989.

\bibitem{Peres:1989:2326}
{A. Peres}.
\newblock {Quantum measurements with postselection}.
\newblock {\em {Phys. Rev. Lett.}}, 62:2326, 1989.

\bibitem{Shabani:07081953}
{A. Shabani and D. A. Lidar}.
\newblock {Linear quantum error correction}.
\newblock e-print arXiv:0708.1953.

\bibitem{Chase:2008:032304}
{B. A. Chase, A. J. Landahl, and J. M. Geremia}.
\newblock {Efficient feedback controllers for continuous-time quantum error
  correction}.
\newblock {\em {Phys. Rev. A.}}, 77:032304, 2008.

\bibitem{Mishra:1977:756}
{B. Mishra and E.C.G Sudarshan}.
\newblock {The Zeno's paradox in quantum theory}.
\newblock {\em {J. Math. Phys.}}, 18:756, 1997.

\bibitem{Ahn:2002:042301}
{C. Ahn, A. C. Doherty, and A. J. Landahl}.
\newblock {Continuous quantum error correction via quantum feedback control}.
\newblock {\em {Phys. Rev. A.}}, 65:042301, 2002.

\bibitem{Ahn:2003:052310}
{C. Ahn, H. W. Wiseman, and G. J. Milburn}.
\newblock {Quantum error correction for continuously detected errors}.
\newblock {\em {Phys. Rev. A.}}, 67:052310, 2003.

\bibitem{Kribs:2006:042329}
{D. W. Kribs, R. W. Spekkens}.
\newblock {Quantum error correcting subsystems are unitarily recoverable
  subsystems}.
\newblock {\em {Phys. Rev. A}}, 74:042329, 2006.

\bibitem{Knill:2006:042301}
{E. Knill}.
\newblock {Protected realizations of quantum infromation}.
\newblock {\em {Phys. Rev. A}}, 74:042301, 2006.

\bibitem{Shibata:1980:891}
{F. Shibata and T. Arimitsu}.
\newblock {Expansion formulas in nonequilibrium statistical mechanics}.
\newblock {\em {J. Phys. Soc. Jpn.}}, 49:891, 1980.

\bibitem{Shibata:1977:171}
{F. Shibata, Y. Takahashi, and N. Hashitsume}.
\newblock {A generalized stochastic liouville equation. Non-Markovian versus
  memoryless master equations}.
\newblock {\em {J. Stat. Phys.}}, 17:171, 1977.

\bibitem{Lindblad:1976:119}
{G. Lindblad}.
\newblock {On the generators of quantum dynamical semigroups}.
\newblock {\em {Comm. Math. Phys.}}, 48:119, 1976.

\bibitem{Krovi:2007:052117}
{H. Krovi, O. Oreshkov, M. Ryazanov, and D. A. Lidar}.
\newblock {Non-Markovian dynamics of a qubit coupled to an Ising spin bath}.
\newblock {\em {Phys. Rev. A}}, 76:052117, 2007.

\bibitem{Wiseman:2006:90}
{H. M. Wiseman and J. F. Ralph}.
\newblock {Reconsidering Rapid Qubit Purification by Feedback}.
\newblock {\em {New J. Phys.}}, 8:90, 2006.

\bibitem{Nakazato:1996:247}
{H. Nakazato, M. Namiki and S. Pascazio}.
\newblock {Temporal behavior of quantum mechanical systems}.
\newblock {\em {Int. J. Mod. Phys. B}}, 10:247, 1996.

\bibitem{Breuer:2002:Oxford}
{H.-P. Breuer and F. Petruccione}.
\newblock {\em {The Theory of Open Quantum Systems}}.
\newblock {Oxford University Press}, {Oxford, UK}, 2002.

\bibitem{Breuer:2004:045323}
{H.-P. Breuer, D. Burgarth and F. Petruccione}.
\newblock {Non-Markovian dynamics in a spin star system: Exact solution and
  approximation techniques}.
\newblock {\em {Phys. Rev. B}}, 70:045323, 2004.

\bibitem{Kerckhoff:0812:1246}
{J. Kerckhoff, L. Bouten, A. Silberfarb, and H. Mabuchi}.
\newblock {Physical model of continuous two-qubit parity measurement in a
  cavity-QED network}.
\newblock e-print arXiv:0812:1246.

\bibitem{Paz:1998:355}
{J. P. Paz and W. H. Zurek}.
\newblock {Continuous error correction}.
\newblock {\em {Proc. R. Soc. London, Ser. A}}, 454:355, 1998.

\bibitem{Jacobs:2004:355}
{K. Jacobs}.
\newblock {Optimal feedback control for the rapid preparation of a single
  qubit}.
\newblock {\em {Proc. of SPIE}}, 5468:355, 2004.

\bibitem{Vaidman:1996:1745}
{L. Vaidman, L. Goldenberg, and S. Wiesner}.
\newblock {Error prevention scheme with four particles }.
\newblock {\em {Phys. Rev. A}}, 54, 1996.

\bibitem{Sarovar:2005:012306}
{M. Sarovar and G. J. Milburn}.
\newblock {Continuous quantum error correction by cooling }.
\newblock {\em {Phys. Rev. A.}}, 72:012306, 2005.

\bibitem{Sarovar:2004:052324}
{M. Sarovar, C. Ahn, K. Jacobs, and G. J. Milburn}.
\newblock {A practical scheme for error control using feedback}.
\newblock {\em {Phys. Rev. A.}}, 69:052324, 2004.

\bibitem{Oreshkov:0812.4682}
{O. Oreshkov}.
\newblock {Topics in quantum information and the theory of open quantum
  systems}.
\newblock Ph.D. thesis, University of Southern California, 2008, e-print
  arXiv:0812.4682.

\bibitem{Oreshkov:2005:110409}
{O. Oreshkov and T. A. Brun}.
\newblock {Weak measurements are universal}.
\newblock {\em {Phys. Rev. Lett.}}, 95:110409, 2005.

\bibitem{Oreshkov:2007:022318}
{O. Oreshkov and T. A. Brun}.
\newblock {Continuous quantum error correction for non-Markovian decoherence}.
\newblock {\em {Phys. Rev. A}}, 76:022318, 2007.

\bibitem{Oreshkov:2008:022333}
{O. Oreshkov, D. A. Lidar, and T. A. Brun}.
\newblock {Operator quantum error correction for continuous dynamics}.
\newblock {\em {Phys. Rev. A}}, 78:022333, 2008.

\bibitem{Blume-Kohout:2008:030501}
{R. Blume-Kohout, H. K. Ng, D. Poulin, and L. Viola}.
\newblock {Constructing qubits in physical systems}.
\newblock {\em {Phys. Rev. Lett.}}, 100:030501, 2008.

\bibitem{Zwanzig:1960:1338}
{R. Zwanzig}.
\newblock {Ensemble method in the theory of irreversibility}.
\newblock {\em {J. Chem. Phys.}}, 33:1338, 1960.

\bibitem{Nakajima:1958:948}
{S. Nakajima}.
\newblock {On Quantum Theory of Transport Phenomena — Steady Diffusion —}.
\newblock {\em {Prog. Theor. Phys.}}, 20:948, 1958.

\bibitem{Brun:2002:719}
{T. A. Brun}.
\newblock {A simple model of quantum trajectories }.
\newblock {\em {Am. J. Phys.}}, 70:719, 2002.

\bibitem{Quang:1997:5238}
{T. Quang, M. Woldeyohannes, S. John, and G. S. Agarwal}.
\newblock {Coherent control of spontaneous emission near a photonic band edge:
  a single-atom optical memory device}.
\newblock {\em {Phys. Rev. Lett.}}, 79:5238, 1997.

\bibitem{Zurek:1984:391}
{W. H. Zurek}.
\newblock {Reversibility and stability of information processing systems}.
\newblock {\em {Phys. Rev. Lett.}}, 53:391, 1984.

\bibitem{Aharonov:1989:2327}
{Y. Aharonov and L. Vaidman}.
\newblock {Aharonov and Vaidman reply}.
\newblock {\em {Phys. Rev. Lett.}}, 62:2327, 1989.

\bibitem{Aharonov:1990:11}
{Y. Aharonov and L. Vaidman}.
\newblock {Properties of a quantum system during the time interval between two
  measurements}.
\newblock {\em {Phys. Rev. A}}, 41:11, 1990.

\bibitem{Aharonov:1988:1351}
{Y. Aharonov, D.Z. Albert, and L. Vaidman}.
\newblock {How the result of a measurement of a component of the spin of a
  spin-$1/2$ particle can turn out to be 100}.
\newblock {\em {Phys. Rev. Lett.}}, 60:1351, 1988.

\end{thebibliography}

\end{document}